\documentclass{aastex}
\usepackage{emulateapj5}
\usepackage{apjfonts}

\def\wig#1{\mathrel{\hbox{\hbox to 0pt{%
          \lower.5ex\hbox{$\sim$}\hss}\raise.4ex\hbox{$#1$}}}}

\shorttitle{Two Classes of Hot Jupiter Atmospheres}
\shortauthors{Fortney}

\newcommand{\mj}{$M_{\mathrm{J}}$}
\newcommand{\rj}{$R_{\mathrm{J}}$}
\newcommand{\me}{$M_{\oplus}$}

\newcommand{\T}{TrES-1}
\newcommand{\hd}{HD 209458b} 
\newcommand{\hh}{HD 149026b}
\newcommand{\he}{HD 189733b}

\newcommand{\te}{$T_{\rm eff}$}
\newcommand{\ti}{$T_{\rm int}$}
\newcommand{\cp}{\citep}
\newcommand{\ct}{\citet}

\begin{document}

\title{A Unified Theory for the Atmospheres of the Hot and Very Hot Jupiters:\\  Two Classes of Irradiated Atmospheres}

\author{J. J. Fortney\altaffilmark{1}$^,$\altaffilmark{2}$^,$\altaffilmark{3}}
\affil{Space Science and Astrobiology Division, NASA Ames Research Center, MS 245-3, Moffett Field, CA 94035}
\email{jfortney@ucolick.org}
\author{K. Lodders}
\affil{Planetary Chemistry Laboratory, Department of Earth and Planetary Sciences, Washington University, St. Louis, MO 63130}

\author{M.~S.~Marley \& R.~S.~Freedman\altaffilmark{2}}
\affil{Space Science and Astrobiology Division, NASA Ames Research Center, MS 245-3, Moffett Field, CA 94035}

\altaffiltext{1}{Spitzer Fellow}
\altaffiltext{2}{Carl Sagan Center, SETI Institute, 515 North Whisman Road, Mountain View, CA 94043}
\altaffiltext{3}{Department of Astronomy and Astrophysics, UCO/Lick Observatory, University of California, Santa Cruz, CA 95064}

\begin{abstract} 

We highlight the potential importance of gaseous TiO and VO opacity on the highly irradiated close-in giant planets.  The atmospheres of these planets naturally fall into two classes that are somewhat analogous to the M- and L-type dwarfs.  Those that are warm enough to have appreciable opacity due to TiO and VO gases we term the ``pM Class'' planets, and those that are cooler, such that Ti and V are predominantly in solid condensates, we term ``pL Class'' planets.  The optical spectra of pL Class planets are dominated by neutral atomic Na and K absorption.  We calculate model atmospheres for these planets, including pressure-temperature profiles, spectra, and characteristic radiative time constants.  Planets that have temperature inversions (hot stratospheres) of $\sim$2000 K and appear ``anomalously'' bright in the mid infrared at secondary eclipse, as was recently found for planets \hh\ and \hd, we term the pM Class.  Molecular bands of TiO, VO, H$_2$O, and CO will be seen in emission, rather than absorption.  This class of planets absorbs incident flux and emits thermal flux from high in their atmospheres.  Consequently, they will have large day/night temperature contrasts and negligible phase shifts between orbital phase and thermal emission light curves, because radiative timescales are much shorter than possible dynamical timescales.  The pL Class planets absorb incident flux deeper in the atmosphere where atmospheric dynamics will more readily redistribute absorbed energy.  This leads to cooler day sides, warmer night sides, and larger phase shifts in thermal emission light curves.  We briefly examine the transit radii for both classes of planets.  The boundary between these classes is particularly dependent on the incident flux from the parent star, and less so on the temperature of the planet's internal adiabat (which depends on mass and age), and surface gravity.  Around a Sun-like primary, for solar composition, this boundary likely occurs at $\sim$0.04-0.05 AU, but uncertainties remain.  We apply these results to pM Class transiting planets that are observable with the \emph{Spitzer Space Telescope}, including \hd, WASP-1b, TrES-3b, TrES-4b, \hh, and others.  The eccentric transiting planets HD 147506b and HD 17156b alternate between the classes during their orbits.  Thermal emission in the optical from pM Class planets is significant red-ward of 400 nm, making these planets attractive targets for optical detection via Kepler, COROT, and from the ground.  The difference in the observed day/night contrast between $\upsilon$ Andromeda b (pM Class) and \he\ (pL Class) is naturally explained in this scenario.

\end{abstract}

\keywords{planetary systems, radiative transfer}

%\newpage

\section{Introduction}
The blanket term ``hot Jupiter" or even the additional term ``very hot Jupiter'' belies the diversity of these highly irradiated planets.  Each planet likely has its own unique atmosphere, interior structure, and accretion history.  The relative amounts of refractory and volatile compounds in a planet will reflect the parent star abundances, nebula temperature, total disk mass, location of the planet's formation within the disk, duration of its formation, and its subsequent migration (if any).  This accretion history will give rise to differences in core masses, total heavy elements abundances, and atmospheric abundance ratios.  Given this incredible complexity, it is worthwhile to first look for physical processes that may be common to groups of planets.

In addition to a mass and radius, one can further characterize a planet by studying its atmosphere.  The visible atmosphere is a window into the composition of a planet and contains clues to its formation history \cp[e.g.,][]{Marley07b}.  Of premier importance in this class of highly irradiated planets is how stellar insolation affects the atmosphere, as this irradiation directly affects the atmospheric structure, temperatures, and chemistry, the planet's cooling and contraction history, and even its stability against evaporation.

Since irradiation is perhaps the most important factor in determining the atmospheric properties of these planets, we examine the insolation levels of the 23 known transiting planets.  We restrict ourselves to those planets more massive than Saturn, and hence for now exclude treatment of the ``hot Neptune'' GJ 436b, which is by far the coolest known transiting planet.  \mbox{Figure~\ref{flux}} illustrates the stellar flux incident upon the planets as a function of both planet mass (\mbox{Figure~\ref{flux}}\emph{a}) and planet surface gravity (\mbox{Figure~\ref{flux}}\emph{b}).  In these plots diamonds indicate transiting planets and triangles indicate other interesting hot Jupiters, for which \emph{Spitzer Space Telescope} data exist, but which do not transit.

The first known transiting planet, \hd, is seen to be fairly representative of these planets in terms of incident flux.  Planets OGLE-TR-56b and OGLE-TR-132b are somewhat separate from the rest of the group because they receive the highest stellar irradiation.  Both orbit their parent stars in less than 2 days and are prototypes of what has been called the class of ``very hot Jupiters'' \cp{Konacki03,Bouchy04} with orbital periods less than 3 days.  However, orbital period is a poor discriminator between ``very hot'' and merely ``hot,'' as \he\ clearly shows.  Labeled a ``very hot Jupiter'' upon its discovery, due to its short 2.2 day period \citep{Bouchy05}, \he\ actually receives a comparatively modest amount of irradiation due to its relatively cool parent star.  Therefore, perhaps a classification based on incident flux, equilibrium temperature, or other attributes would be more appropriate.  In this paper we argue that based on the examination of few physical processes that two classes of hot Jupiter atmospheres emerge with dramatically different spectra and day/night contrasts.  Equilibrium chemistry, the depth to which incident flux will penetrate into a planet's atmosphere, and the radiative time constant as a function of pressure and temperature in the atmosphere all naturally define two classes these irradiated planets.

Our work naturally builds on the previous work of \ct{Hubeny03} who first investigated the effects of TiO and VO opacity on close-in giant planet atmospheres as a function of stellar irradiation.  These authors computed optical and near infrared spectra of models with and without TiO/VO opacity.  In general they found that models with TiO/VO opacity feature temperature inversions and molecular bands are seen in emission, rather than absorption.  Two key questions from the initial \ct{Hubeny03} investigation were addressed but could not be definitely answered were: 1) if a relatively cold planetary interior would lead to Ti/V condensing out deep in the atmosphere regardless of incident flux, thereby removing gaseous TiO and VO, and, 2) if this condensation did not occur, at what irradiation level would TiO/VO indeed be lost at the lower atmospheric temperatures found at smaller incident fluxes.

Later \ct{Fortney06} investigated model atmospheres of planet \hh\ including TiO/VO opacity at various metallicities.  Particular attention was paid to the temperature of the deep atmosphere \emph{P-T} profiles (as derived from an evolution model) in relation to the Ti/V condensation boundary.  Similar to \ct{Hubeny03}, they found a temperature inversion due to absorption by TiO/VO and computed near and mid-infrared spectra that featured emission bands.  Using the \emph{Spitzer} InfraRed Array Camera (IRAC) \ct{Harrington07} observed \hh\ in secondary eclipse with \emph{Spitzer} at 8 $\mu$m and derived a planet-to-star flux ratio consistent with a \ct{Fortney06} model with a temperature inversion due to TiO/VO opacity.  At that point, looking at the work of \ct{Fortney06} and especially \ct{Hubeny03}, \ct{Harrington07} could have postulated that all objects more irradiated than \hh\ may possess inversions due to TiO/VO opacity, but given the single-band detection of \hh, caution was in order.  More recently, based on the four-band detection of flux from \hd\ by \ct{Knutson08}, \ct{Burrows07c} find that a temperature inversion, potentialy due to TiO/VO opacity, is necessary to explain this planet's mid-infrared photometric data.  Based on their new \hd\ model and the previous modeling investigations these authors posit that planets warmer than \hd\ may features inversions, while less irradiated objects such as \he\ do not, and discuss that photochemical products and gaseous TiO/VO are potential absorbers which may lead to this dichotomy.

We find, as has been previously shown, that those planets that are warmer than required for condensation of titanium (Ti) and vanadium (V)-bearing compounds will possess a temperature inversion at low pressure due to absorption of incident flux by TiO and VO, and will appear ``anomalously'' bright in secondary eclipse at mid-infrared wavelengths.  Thermal emission in the optical will be significant \cp{Hubeny03,Lopez07}.  Furthermore, here we propose that these planets will have large day/night effective temperature contrasts.  We will term these very hot Jupiters the ``pM Class,'' meaning gaseous TiO and VO are the prominent absorbers of optical flux.  The predictions of equilibrium chemistry for these atmospheres are similar to dM stars, where absorption by TiO, VO, H$_2$O, and CO is prominent \cp{Lodders02b}.  Planets with temperatures below the condensation curve of Ti and V bearing compounds will have a gradually smaller mixing ratio of TiO and VO, leaving Na and K as the major optical opacity sources \cp*{BMS}, along with H$_2$O, and CO.  We will term these planets the pL class, similar to the dL class of ultracool dwarfs.  These planets will have relatively smaller secondary eclipse depths in the mid infrared and significantly smaller day/night effective temperature contrasts.  As discussed below, published \emph{Spitzer} data are consistent with this picture.  The boundary between these classes, at irradiation levels (and atmospheric temperatures) where Ti and V may be partially condensed is not yet well defined.

In this paper we begin by discussing the observations to date.  We then give an overview of our modeling methods and the predicted chemistry of Ti and V.  We calculate pressure-temperature (\emph{P-T}) profiles and spectra for models planets.  For these model atmospheres we then analyze in detail the deposition of incident stellar flux and the emission of thermal flux, and go on to calculate characteristic radiative time constants for these atmospheres.  We briefly examine transmission spectra before we apply our models to known highly irradiated giant planets.  Before our discussion and conclusions we address issues of planetary classification.

\section{Review of Spitzer Observations}
\subsection{Secondary Eclipses}
The \emph{Spitzer Space Telescope} allows astronomers to measure the thermal emission from $\sim$3-30 $\mu$m from the highly irradiated atmospheres of these EGPs.  This field has progressed quickly from the first observations of secondary eclipses (when the planet passes behind its parent star) by \ct{Charb05} for \T\ and by \ct{Deming05b} for \hd.  Once it was clear what \emph{Spitzer}'s capabilities were for these planets, additional observations came quickly.  These included secondary eclipses for \he\ \cp{Deming06}, \hh\ \cp{Harrington07}, GJ 436b \cp{Deming07,Demory07}, and a wealth of new (yet to be published) data for $\sim$10 other systems.  In addition, mid infrared spectra were obtained for \hd\ \cp{Richardson07,Swain07} and \he\ \cp{Grillmair07}.  In particular, \ct{Richardson07} claim evidence for emission features in the mid-infrared spectrum.  Very recently \ct{Knutson08} observed the secondary eclipse of \hd\ in all four InfraRed Array Camera (IRAC) bands.

One can now begin to examine the secondary eclipse data for trends in planetary brightness temperatures.  The brightness temperature, $T_{\rm B}$ is defined as the temperature necessary for a blackbody planet to emit the flux observed in the given observational wavelength or band \cp{ChambHunt}.  For the published secondary eclipse data as of June 2007, \ct{Harrington07} show that the observed day side brightness temperatures for most transiting planets (\T, \he, \hd) vary by $\pm20$\% from calculated planetary equilibrium temperatures (when making a uniform assumption for all planets regarding albedo and flux redistribution).  These differences are presumably due to wavelength dependent opacity and varying dynamical redistribution of the absorbed flux.  However, \hh\ has a $\sim$45\% higher brightness temperature than predicted from this relation, meaning that the temperature one measures at 8 $\mu$m is significantly higher than the planet's equilibrium temperature.  This level of infrared emission was predicted by \ct{Fortney06} for models of \hh\ that included a hot stratosphere induced by absorption of stellar flux by TiO and VO.  More recently, the \ct{Knutson08} observations of \hd\ are broadly similar to the 1-band detection for \hh.  Relatively high brightness temperatures of 1500-1900 K were observed from 3.6 to 8 $\mu$m, compared to the \ct{Harrington07} predicted equilibrium temperature of 1330 K.  These observations have also been interpreted as being caused by a temperature inversion (hot stratosphere) by \ct{Burrows07c}.  As we will discuss in \S5.1, \hd\ is in a temperature regime that is difficult to model, where Ti and V are expected to be partially condensed.

\subsection{Phase Curves}
The observations that best allow us to understand the atmospheric dynamics of these presumably tidally locked planets are those made as a function of orbital phase.  Two approaches have been used to measure the thermal emission of these planets away from secondary eclipse.  The first is to periodically observe the infrared flux from the combined planet+star system at several times in the planet's orbit.  The advantage of periodic short observations is that little telescope time is used.  The disadvantage is that since the system is not monitored continuously, systematic uncertainties could be important and may be hard to correct for.  There are published data for four planets that have been observed in this way: detections of phase variation for $\upsilon$ And b \cp{Harrington06} and HD 179949b \cp{Cowan07}, and upper limits for \hd, 51 Peg b \cp{Cowan07}.  Alternatively, one could monitor a system continuously over a significant fraction of a planet's orbital period, to eliminate any uncertainties induced by having to revisit the target.  This method was employed by \ct{Knutson07b} to observe \he\ continuously for 33 hours, including the transit and secondary eclipse.  Large day/night contrasts were found for pM Class planets $\upsilon$ And b (which is consistent with a dark night side) and HD 179949b, whereas pL Class planet \he\ has a day/night 8 $\mu$m brightness temperature difference of only $\sim$240 K.  We will return to the \ct{Cowan07} upper limit for \hd\ in \S8.

Two additional factors that affect these observations are the brightness of the parent star and the inclination  of the planetary orbit.  While the brightest planet-hosting stars allow for the largest flux measurements (e.g., $\upsilon$ And b), the brightest hot Jupiter systems do not transit.  A measurement of the secondary eclipse depth while obtaining a light curve is extremely valuable, because the planet-to-star flux ratio at full planet illumination is thus known, as is the planet's radius from transit observations.  The interpretation is then much more straightforward for the transiting systems.
 
Among the current published data there appear to be two planets that are bright in secondary eclipse, \hh\ and \hd, compared to other transiting planets, and two planets that appears to have large day/night temperature differences, $\upsilon$ And b and HD 179949b.  We will show that  these planets, \hh, \hd, $\upsilon$ And b, and HD 179949b, are ``pM Class'' planets, while the less irradiated hot Jupiters (such as \he\ and \T) are ``pL Class.''  We will now turn to models of these classes of planets to examine how and why their atmospheres differ so strikingly.

\section{Model Atmospheres}
\subsection{Methods}
We have computed atmospheric pressure-temperature (\emph{P--T}) profiles and spectra for several planets with a plane-parallel model atmosphere code that has been used for a variety of planetary and substellar objects.  The code was first used to generate profiles and spectra for Titan's atmosphere by \citet{Mckay89}.  It was significantly revised to model the atmospheres of brown dwarfs \citep{Marley96, Burrows97, Marley02} and irradiated giant planets \citep[][for Uranus]{MM99}.  Recently it has been applied to L- and T-type brown dwarfs \citep{Saumon06,Saumon07,Cushing08} and hot Jupiters \citep{Fortney05,Fortney06,Fortney07b}.  It explicitly accounts for both incident radiation from the parent star and thermal radiation from the planet's atmosphere and interior.  The radiative transfer solution algorithm was developed by \citet{Toon89}.  We model the impinging stellar flux from 0.26 to 6.0 $\mu$m and the emitted thermal flux from 0.26 to 325 $\mu$m.

We use the elemental abundance data of \citet{Lodders03} and chemical equilibrium compositions are computed with the CONDOR code, following \citet{Lodders02,Lodders06}, and \citet{Lodders99,Lodders02b}.  We maintain a large and constantly updated opacity database that is described in \ct{Freedman07}.  When including the opacity of clouds, such as Fe-metal and Mg-silicates, we use the cloud model of \citet{AM01}.  However, in our past work we have found only weak effects on \emph{P-T} profiles and spectra due to cloud opacity \cp{Fortney05}, so we ignore cloud opacity here.  However, the sequestering of elements into condensates, and their removal from the gas phase (``rainout'') is always accounted for in the chemistry calculations.  We note that day sides of the strongly irradiated pM Class planets are too warm for Fe-metal and Mg-silicates condensates to form.

\subsection{Very Hot Jupiters and TiO/VO Chemistry}
It is clear that the abundances to TiO and VO gases is important in these atmospheres.  \ct{Hubeny03} first discussed how understanding the ``cold trap'' phenomenon may be significant in understanding these abundances.  If a given \emph{P-T} profile crosses a condensation curve in two corresponding altitude levels, the condensed species is expected to eventually mix down to the highest pressure condensation point, where the cloud remains confined due to the planet's gravitational field.  This process is responsible for the extremely low water abundance in the Earth's stratosphere.  It can also be seen in the atmospheres Jupiter and Saturn, where the ammonia ice clouds are confined to a pressure of several bars, although both these planets exhibit stratospheres, such that their warm upper atmospheres pass the ammonia condensation curve again at millibar pressures.  For the highly irradiated planets, the relevant condensates are those that remove gaseous TiO and VO, and sequester Ti and V into solid condensates at pressures of tens to hundreds of bars, far below the visible atmosphere.

The cold trap phenomenon constitutes a departure from chemical equilibrium that cannot be easily accounted for in the pre-tabulated chemical equilibrium abundances used by \ct{Hubeny03}, \ct{Fortney06}, \ct{Burrows07c}, and here as well.  Our chemical abundances and opacities are pre-tabulated in \emph{P-T} space and the atmosphere code interpolates in these abundances as it converges to a solution.  The abundances determined for any one pressure level of the \emph{P-T} profile are not cognizant of abundances of other levels of the profile, although condensation and settling of species is always properly accounted for.  In this case a tabulated TiO abundance at a given \emph{P-T} point at which, in equilibrium, TiO would be in the gaseous state (warmer than required for Ti condensation), may not be correct.  If the atmospheric \emph{P-T} profile intersects the condensation curve, the atmosphere becomes depleted in TiO above the cloud.  We do not treat the cold trap here.  In practice, we use two different opacity databases:  one with TiO and VO removed at P$<$10 bars, which simulates the removal of Ti and V into clouds, and one in which gases TiO and VO remain as calculated by equilibrium chemistry (which does not include depletion from a cold trap).

A full discussion of titanium and vanadium chemistry in the context of M- and L-dwarf atmospheres can be found in \citet{Lodders02b}.  Much of that discussion pertains to the atmospheres of highly irradiated planets as well.  The chemistry is complex.  For instance \citet{Lodders06} (using the updated Lodders, 2003 abundances) find the first Ti condensate will not necessarily be CaTiO$_3$.  For solar metallicity the first condensate is TiN if $P \gtrsim 30$ bars, Ca$_3$Ti$_2$O$_7$ if $5 \lesssim P \lesssim 30$ bars, Ca$_4$Ti$_3$O$_{10}$ if $0.03 \lesssim P \lesssim 5$ bars, and CaTiO$_3$ if $P \lesssim 0.03$ bars.  These four condensates are the \emph{initial} condensates as a function of total pressure and their condensation temperatures define the Ti-condensation curve in \mbox{Figure~\ref{pt1}}.  Another important point is that, following \citet{Lodders02b} we assume that vanadium condenses into solid solution with Ti-bearing condensates, as is found in meteorites \cp{Kornacki86}.  \citet{Lodders02b} find this condensation sequence is also consistent with observed spectra at the M- to L-dwarf transition \citep{Kirkpatrick99}.  In the absence of this effect V would not condense until $\sim$200 K cooler temperatures are reached and solid VO forms, as in the chemistry calculations of \ct{Burrows99}, \ct{Allard01}, and \ct{Sharp07}.  Choices made in the calculation of the Ti condensation curve (e.g., number of potential Ti-bearing compounds included in the calculations) and the V condensation curve (e.g.~whether V enters into Ti-bearing condensates or condenses as solid VO) will cause some shift in the position of these condensation curves, important boundaries between the pL and pM Classes.

\subsection{Calculating Pressure-Temperature Profiles}
When modeling the atmospheres of irradiated objects with a one-dimensional, plane parallel atmosphere code it is necessary to weigh the stellar flux by a geometric factor if one is computing a profile for day-side average or planet-wide average conditions.  Decriptions of this issue in solar system atmosphere modeling can be found in \ct{Appleby84} and \ct{Mckay89}.  In the context of highly irradiated EGPs, \citet{Burrows04,Burrows06,Burrows07}, \ct{Barman05}, \ct{Seager05}, and \ct{Fortney07b} have all discussed this issue in some detail.  All are approximations for atmospheric structures that are surely complex for these highly irradiated atmospheres.  In this paper, we make a straightforward choice to multiply the incident flux by a factor, $f$=0.5 for a day-side average (since the atmosphere radiates over 2$\pi$ steradians but intercepts flux over only $\pi$ radians), and assume normal incidence of flux, $\mu$=1.  This differs slightly from choices in our previous papers \cp[see][]{Fortney07b}.  All profiles shown in this paper are relevant for the irradiated planetary day side.

In \mbox{Figure~\ref{pt1}} we show \emph{P-T} profiles as a function of distance from the Sun.  At 0.025 and 0.03 AU the profiles are everywhere warmer that the condensation curve for a solar abundance of Ti and V.  Essentially all Ti and V is expected to be in TiO and VO.  By 0.035 AU, where the incident flux is weaker, the profile crosses this condensation curve, labeled ``1X Ti/V-Cond."  To the left of this curve, the TiO/VO abundances fall exponentially with decreasing temperature.  To the right, the TiO/VO abundances are essentially constant.  This particular \emph{P-T} profile is oddly shaped since it samples a region of \emph{P-T} space where the opacity drops rapidly with temperature.  It crosses back to the right of the condensation curve, where it again finds the ``full'' abundance of TiO/VO.  This is an example of a profile that violates the cold-trap (which we do not model), as at lower pressures TiO/VO abundances should be reduced, given the condensation of Ti and V below.  The dotted curve labeled ``90\% Ti-Cond'' shows where 90\% of Ti has been lost to a solid condensate  (Lee \& Lodders, in prep).  \mbox{Figure~\ref{opac}} is a plot of optical and near infrared opacity at 1 mbar and shows that TiO/VO are still major optical opacity sources even after condensation has begun.  At 1700 K, before Ti/V condensation, opacity (here $\log(\kappa)$ in cm$^2$g$^{-1}$) in V-band is $\sim$-0.5.  By 1600 K, 33 K cooler than the ``90\% Ti-Cond'' curve, this opacity is still an order of magnitude larger than the opacity at 1400 K, by which point the strong TiO/VO bands have given way to alkali lines and water bands.

That TiO/VO gradually wane in importance is consistent with observations of dwarfs at the M-to-L transition \cp{Lodders02b}.  The mixing ratios of these species do not drop to zero at the solar abundances condensation curve.  For model atmospheres more distant from the parent star (0.05 AU and beyond), the abundance of TiO/VO continues to fall exponentially with temperature, such that in the upper atmosphere it is no longer a major opacity source.

We can examine how the derived \emph{P-T} profiles for the pM Class planets vary as a function of $T_{\rm int}$, a planet's intrinsic effective temperature in the absence of irradiation, and as a function of surface gravity.  This is shown in \mbox{Figure~\ref{pt2}}.  Lower gravity planets have lower pressure photospheres, and consequently have warmer interiors, all else being equal, which has long been known from basic theory.  As can be seen in \mbox{Figure~\ref{flux}}\emph{b}, the lowest gravity planets cluster around $g$=9 m s$^{-2}$, while the highest gravity planets, with the exception of HD 147506b, cluster around $g$=25 m s$^{-2}$, similar to Jupiter.  Both of these values are a factor of 2/3 removed from our baseline $g$=15 m s$^{-2}$.  The effect of gravity on the \emph{P-T} profiles for the pM Class planets are relatively modest.  However, the effect of \ti\ is potentially important.  We show profiles at \ti\ values of 150, 200, and 250 K.  Generally for adiabatic interiors, the higher the \ti, the warmer the interior, and the larger the radius of the planet.  Independent of \ti, which could in principle be affected by barriers to convection \cp{Stevenson85, Chabrier07c}, the higher the entropy of the interior adiabat, the warmer the atmospheric adiabat, the larger the radius of the planet, and the less likely the deep atmosphere profile at 10 to 1000 bar will cross the Ti/V condensation boundary.  Over Gyr-timescales it may be possible for a planet's interior to cool enough such that significant Ti/V condensation could occur at depth, and a planet could transition from pM to pL Class.

In \mbox{Figure~\ref{flux}} the gray shaded area is meant to illustrate a transition region between these classes of planets, where TiO and VO are waning in importance.  Given the lack of data for the systems in this proposed transition region, the exact location and height of this region is merely suggestive.  Analysis of the current secondary eclipse data for \he\ \cp{Deming06, Grillmair07, Knutson07b} fits well with a pL Class model \cp{Fortney07b}, and \ct{Knutson07b} directly measured a small day-night contrast, suggesting that TiO/VO opacity has waned considerably at this incident flux level and there is no temperature inversion.  We next turn to the upper boundary.

At the high incident fluxes encountered by \hh, the temperature inversion predicted by \ct{Fortney06} neatly explains the large planet-to-star flux ratio at 8 $\mu$m \cp{Harrington07}.  At lower incident fluxes, the large day-night temperature contrasts implied for $\upsilon$ And b and HD 179949 imply temperature inversions as we'll show in \S4.2 (and pM classification) at this flux level.  Intriguingly, at a still lower irradiation level, \ct{Burrows07c} find that a strong temperature inversion is needed for \hd, which we also address in \S5.1, which indicates that inversions extend down to at least this incident fluxs level.  This inversion persists even though it is clear from \mbox{Figure~\ref{pt1}} that Ti/V condensation has begun in our 1D model at this irradiation level (although is has not for the more irradiated $\upsilon$ And b, as we show in \S5.2).  Similar to the findings of \ct{Burrows07c}, we can only obtain a temperature inversion for \hd\ induced by TiO/VO opacity if their abundances are not diminished by the simple cold trap description discussed above.  In \S5.1 and \S7.1 we discuss a few avenues to towards understanding if TiO/VO remains the prominent stratospheric absorber down to this incident flux level.

The transition area as plotted in \mbox{Figure~\ref{flux}} is thus given from mix of the observations and theory available to date.  Additional complications will be effects of atmospheric metallicity on Ti/V condensation, the treatment of the incident stellar flux, planet surface gravity, and the temperature of the interior adiabat.  For instance, it is certainly possible that a planet could appear marginally on to the hot side of this divide, but with a very cool interior (due to low mass and/or old age) would have Ti/V condensed such that it is a pL Class member.  With better knowledge of the incident flux deposition and Ti/V condensation, it may eventually be possible to ``take the temperature of the interior adiabat'' at pressures of 10 to 100 bars, which could shed light on the interior structure of some of these planets.

\section{Results and Predictions}
\subsection{Absorption and Emission of Flux}
It is worthwhile to examine in some detail how the atmospheres of the pM Class and pL Class planets differ in the absorption of stellar flux as a function of atmospheric pressure.  One can plot the layer net fluxes for several pressures as a function of wavelength to illustrate the main atomic and molecular absorbers in the atmosphere.  Since these models are in radiative equilibrium, the integrated absorbed flux within a layer should equal the integrated emitted flux from that same layer.  We can plot the net and integrated thermal flux to examine at what wavelengths, and which species, lead to atmospheric heating and cooling.  A series of plots are shown in \mbox{Figures~\ref{v5}} and \ref{i5}, for a pL Class planet at 0.05 AU, and \mbox{Figures~\ref{v3}} and \ref{i3}, for a pM Class planet at 0.03 AU.  The left ordinate of these plots shows the quantity erg s$^{-1}$ g$^{-1}$ $\mu$m$^{-1}$, which is the power per wavelength interval, per gram of atmosphere, absorbed or emitted in a slab of atmosphere, though we'll maintain the term ``flux'' for convenience.

In \mbox{Figure~\ref{v5}} five panels are shown, from 0.45 mbar to 4.1 bar, each separated by approximately an order of magnitude in pressure.  Prominent in the top two panels are absorption of incident flux by neutral atomic alkalis Na (at 0.59 $\mu$m) and K (at 0.77 $\mu$m) as well as strong H$_2$O bands in the near infrared.  While Na and K lead to strong absorption in optical wavelengths, causing extremely dark atmospheres in the optical \cp{Sudar00,SWS00}, at 0.45 and 4.4 mbar, respectively, 4 and 1.5 times more incident stellar is actually absorbed by H$_2$O than by alkalis.  The third panel, 43 mbar, shows that essentially all incident flux at the wavelengths of the Na and K line cores have been absorbed, and the alkali line wings are now as important as the H$_2$O bands in absorption.  By 420 mbar, the wavelengths corresponding to the broad Na and K lines have already become optically thick, leaving little flux left.  About $\sim$10$^3$ less flux is absorbed at 4.1 bar than 0.45 mbar.  The pressure is now in the nearly isothermal region of the \emph{P-T} profile, as shown in \mbox{Figure~\ref{pt1}}.  In summary, Na, K, and H$_2$O are the major absorbers and heating in the millibar region of the atmosphere amounts to $\sim10^6$ erg g$^{-1}$ s$^{-1}$.

We can examine \mbox{Figure~\ref{i5}} to understand the cooling of the atmosphere as these same pressure levels.  This figure has a much larger wavelength range that is shown on a log scale.  In the top three panels, from 0.45 to 43 mbar, it is predominantly the near and mid infrared water bands that allow the atmosphere to cool to space. Between $\sim$0.8-1.8 $\mu$m, positive layer net fluxes occur due to absorption of flux from above, by H$_2$O.  Deeper in the atmosphere, at higher temperatures, the local Planck function at a given layer overlaps shorter wavelength water bands from $\sim$1-2 $\mu$m that are primarily responsible for cooling.

In \mbox{Figure~\ref{v3}} we can clearly see how this pM Class planet differs from the pL Class.  The incident stellar flux is higher and more of this flux is deposited into the 0.45 mbar layer of this model, as compared to that shown in \mbox{Figure~\ref{v5}}.  This deposition occurs almost entirely through absorption by TiO and VO bands in the optical and near infrared.  Water vapor still absorbs flux, but its effect is swamped by the TiO/VO absorption.  This continues at 4.4 mbar, but already by 43 mbar, the vast majority of the incident flux has been absorbed, especially in the optical wavelengths.  By 420 mbar there is little flux left in the optical and by 4.1 bar there is little flux left at any wavelength; again this pressure is in the nearly isothermal region shown in \mbox{Figure~\ref{pt1}}.

The top panel of \mbox{Figure~\ref{i3}} shows that, since the atmosphere has absorbed a significant amount of flux at low pressures, compared to the pL Class model, that the atmosphere must reach a much higher temperature to be able to adequately radiate away this energy.  The local Planck function moves significantly blue-ward, so that it can radiate by means of the strong H$_2$O near infrared bands as well as the optical bands of TiO and VO.  At 4.4 and 43 mbar, the atmosphere is cooler, and radiation by H$_2$O bands is more important.  At 43 mbar some heating from above does occur via TiO/VO absorption.  At 420 mbar and 4.1 bar, the local temperature has again increased, leading to radiation again most prominently by the near infrared H$_2$O bands.

\subsection{Radiative Time Constants}
It is clear that a key difference between the atmospheres of the pL Class planets and pM Class planets is the pressures at which the absorption and emission of flux occurs.  This can be shown in a more straightforward manner by means of the brightness temperature, $T_{\rm B}$.  One can then examine the \emph{P-T} profile to find the pressure that corresponds to a given $T_{\rm B}$, which is the characteristic atmospheric pressure for this thermal emission.  This quantity is plotted as a function of wavelength in \mbox{Figure~\ref{ptau}}.  In general thermal emission arises from a pressure level roughly an order of magnitude greater in a pL Class atmosphere than in a pM Class atmosphere, due to the higher opacity in the hotter pM atmospheres, both at optical wavelengths, due to TiO/VO, and infrared wavelengths, due to increased H$_2$O opacity and H$_2$ collision induced absorption.

What effect this may have on the dynamical redistribution of energy in a planetary atmosphere can be considered after calculation of the radiative time constant, $\tau_{\rm rad}$.  In the Newtonian cooling approximation a temperature disturbance relaxes exponentially toward radiative equilibrium with a characteristic time constant $\tau_{\rm rad}$ \cp[e.g.,][]{GoodyYung,Salby}.  At photospheric pressures this value can be approximated by
\begin{equation}
\label{trad1}
\tau_{\rm rad} \sim \frac{P}{g} \frac{c_{\rm P}}{4 \sigma T^3},
\end{equation}
where $\sigma$ is the Stefan-Boltzmann constant and $c_{\rm P}$ is the specific heat capacity \cp{Showman02}.  However, this quantity can be derived anywhere in the radiative atmosphere via the following formulation.  We first obtain a \emph{P-T} profile solution that is in radiative-convective equilibrium.  We can then take this profile, and in a given layer include a small ($\sim$10 K) thermal perturbation, $\Delta T$.  With this $\Delta T$ in place we perform a radiative transfer calculation, and additionally calculate the flux divergence $dF/dz$, which is zero in radiative equilibrium (where $z$ is the height).  For a function $f(t)$ a time constant will be given by $f \times (df/dt)^{-1}$.  We can then solve for $\tau_{\rm rad}$ as
\begin{equation}
\label{trad2}
\tau_{\rm rad} = \Delta T \frac{\rho c_{\rm P}}{dF/dz},
\end{equation}
where $\rho$ is the mass density.  The quantity $(dF/dz)/(\rho c_{\rm P})$ is the heating/cooling rate in K s$^{-1}$.  By varying the location of the $\Delta T$ perturbation with height, one can calculate $\tau_{\rm rad}$ as a function of pressure in the atmosphere, for a given \emph{P-T} profile.  In practice we find values of $\tau_{\rm rad}$ that are very similar to those derived by \ct{Iro05} for \hd.  Interestingly, these authors performed a much different calculation that involved the input of Gaussian temperature fluctuations and the direct calculation of the time necessary for their code to relax back to radiative equilibrium.  A larger exploration of $\tau_{\rm rad}$ over a wider range of phase space, with additional discussion, will be found in A.~P.~Showman et al. (in prep.).

In \mbox{Figure~\ref{PTtau}} we show our calculated $\tau_{\rm rad}$ as a function of pressure alongside the radiative-convective equilibrium profiles from which they were generated.  Although these time constants are nearly equivalent in the dense lower atmosphere, they are significantly different in the thinner upper atmosphere that one is sensitive to from mid-infrared observations.

The right y-axis in \mbox{Figure~\ref{ptau}} is cast in terms of the radiative time constant appropriate for a given pressure in the atmosphere.  Note that this axis is not linear, but each major tick mark is labeled with the $\tau_{\rm rad}$ that is appropriate for the major tick mark at the left.  While those for the pM Class range from only $10^3-10^4$ s, those for the pL Class range from $10^4-10^5$ seconds.  This $\sim 10 \times$ difference in timescale is a consequence of both the hotter temperatures and lower pressures of the pM Class photospheres, as can be understood from \mbox{Equation~\ref{trad1}}.

One can also define an advective time scale, a characteristic time for moving atmospheric gas a given planetary distance.  A common definition is
\begin{equation}
\label{trad3}
\tau_{\rm advec} = \frac{R_{\rm p}}{U},
\end{equation}
where $U$ is the wind speed and $R_{\rm p}$ is the planet radius \cp{Showman02,Seager05}.  If one sets $\tau_{\rm advec}=\tau_{\rm rad}$ and sets $R_{\rm p}=1 R_{\rm J}$, one can derive the wind speed $U$ that would be necessary to advect atmospheric gas before a $\tau_{\rm rad}$ has elapsed.  This is also shown in \mbox{Figure~\ref{ptau}}, via the gray bars on the right side.  Since depth to which one ``sees'' is wavelength dependent, the apparent efficiency of atmospheric dynamics in redistributing energy in the planet's atmosphere will depend on the wavelength of observations.  For wavelengths where the atmospheric opacity is low, one sees flux emerging from relatively deep layers.  The planet will appear more homogenized \cp{Seager05}, as was shown explicitly by \ct{Fortney06b} for light curves for the \ct{CS06} 3D dynamical model.  Therefore the shape and times of maxima and minima in thermal emission light curves will be a function of wavelength.

Examination of \mbox{Figure~\ref{ptau}} shows that for a pM Class planet winds speeds of \emph{dozens} of km s$^{-1}$ would be necessary for advection to dominate over cooling/heating!  For a pL Class planet more modest winds speeds of several km s$^{-1}$ are needed.  Several km s$^{-1}$ winds, which are similar to the sound speed, are in line with predictions of a number of 2D and 3D dynamical models for hot Jupiter atmospheres \cp{Showman02,CS05,Langton07,Dobbs07}.  These calculated wind speeds are far below those needed to advect air before it radiatively heats/cools in a pM Class atmosphere.  

Winds will not be able to redistribute energy at the photospheres of pM Class planets.  The atmospheres of pM Class planets likely appear as one would expect from radiative equilibrium considerations:  the hottest part of the atmosphere is at the substellar point, and the atmosphere is cooler as one moves toward the planet's limb.  Deviations from this prediction may occur from 1-2 $\mu$m where opacity windows allow for observation of flux from relatively higher pressures.  The night-side temperature will be relatively cold and will be set by the efficiency of energy redistribution at depth, as well as the intrinsic flux from the interior of the planet.  In the absence of a large intrinsic flux driven by an interior energy source, such as tides, redistribution may well swamp the intrinsic flux.    For a pL Class atmosphere, there will be a much more complex interplay between radiation and dynamics; energy redistribution will more readily lead to a planetary hot spot being blown downwind.  The location of the hot spot will itself be wavelength dependent, and there will increased energy transfer to the night side at photospheric pressures.  Recently \ct{Dobbs07} have shown that larger day-night temperature differences are expected with increased atmospheric opacity, but they models lack motivation for their opacity choices.  Dynamical models that do not include a realistic treatment of opacities and radiative transfer will miss the important differences between these two classes of planets.

\subsection{Characterization Via Transit Observations}
Primary transit observations for pL Class planets \hd\ and \he\ have yielded detections of sodium \cp{Charb02} and water vapor \cp{Barman07,Tinetti07}.  Since water vapor is present in abundance in pM Class planets, it should be detectable as well.  Observations in the optical would show strong absorption due to TiO and VO across a broad wavelength range, which will lead to observed transit radii \cp{Hubbard01,Baraffe03,Burrows03} that correspond to even lower atmospheric pressure (higher in the atmosphere) that those for pL Class planets.

We perform a simple calculation to illustrate this point.  From \mbox{Figure~\ref{ptau}} the approximate pressure levels for thermal emission are known.  After calculating the distances between the wavelength-dependent photospheric pressures to a reference pressure of 1 bar, we can follow the implementation of \ct{Burrows07} to derive the (even larger) radius at which the \emph{slant optical depth} reaches unity.  This is the radius one measures (at a given wavelength or waveband) during a transit observation.  From \ct{Burrows07}, the extension $dR$ from the photospheric radius to the transit radius is approximately
\begin{equation}
\label{dr}
dR=H \ln \sqrt{\frac{2\pi R_{\mathrm p}}{H}},
\end{equation}
where $H$ is the characteristic atmospheric scale height and $R_{\rm p}$ is the planet radius.  We plot the approximate transit radius for the model pL Class planet and pM Class planet in \mbox{Figure~\ref{rad}}.  Here the 1 bar radius is 1.20 \rj\ and the characteristic $T$ for the scale height calculation for the pL and pM Class planets are 1070 K and 2000 K, respectively.  Since our derived \emph{P-T} profiles do not reach to microbar pressures, a more detailed transit radius calculation utilizing the precise run of temperature and opacity vs.~pressure \cp[see][]{Hubbard01,Fortney03} will be investigated in future work.  Nonetheless, the important points are clear.  Due predominantly to the hotter atmosphere of pM Class planets, their atmospheric extension is larger.  Additionally, as suggested by \ct{Burrows07c} the strong opacity of TiO and VO leads to a larger radius in a wide optical band, compared pL Class planets.  \mbox{Figure~\ref{rad}} shows a difference of $\sim$5\% at optical wavelengths.  This difference in radius could be even further enhanced by temperatures in excess of the 2000 K used in the calculation.  This plot should only be considered suggestive, as a more detailed calculation would involve different chemical abundances on the day and night side of the planetary limb.  Cooler limb temperatures may not allow for gaseous TiO and VO.

Detailed secondary eclipse observations may shed additional light on the structure of these atmospheres.  As was shown in \ct{Fortney06} for \hh, planets with temperature inversions will show limb brightening, rather than darkening, which may eventually be detectable via secondary eclipse mapping \cp{Williams06,Rauscher07b}.

\section{Application to Known Planets}
\subsection{HD 209458b at Secondary Eclipse}
Recently \citet{Knutson08} published observations of the secondary eclipse of planet \hd.  These observations utilized \emph{Spitzer} IRAC, and constitute the first published observations for a transiting planet across all four IRAC bands.  These observed planet-to-star flux ratios were interpreted as being caused by a temperature inversion (hot stratosphere) on the day side of the planet \cp{Burrows07c,Knutson08}.  The reasons for this interpretation include a large brightness temperatures relative to its expected day-side \te, especially at 4.5 and 5.8 $\mu$m, and a large flux ratio at 4.5 $\mu$m (relative to 3.6 $\mu$m), which is enhanced due to strong CO and H$_2$O emission features from 4-6 $\mu$m \cp{Fortney06,Burrows07c}.  In our terminology, \hd\ joined \hh\ as a pM class planet.

In \mbox{Figure~\ref{209}}\emph{a} we plot \emph{P-T} profiles for \hd, with and without the opacity of TiO and VO.  As was previously shown \citep{Hubeny03,Fortney06,Burrows07c} absorption of stellar flux by TiO/VO leads to a hotter upper atmosphere and cooler lower atmosphere, for a given planet.  The profile that includes TiO/VO opacity (red) is cooler than the condensation curve for a solar-abundance mixture of Ti/V.  For this profile, given our equilibrium chemistry table, the abundance of TiO gas drops even below 10\% of a solar Ti abundance, but then increases back to solar in the hot upper atmosphere.  A proper treatment of the cold trap could leave only a fraction of a solar abundance of Ti/V in TiO/VO in the upper atmosphere.  That being said, a hot stratosphere similar to one that we derive is nessary to explain the planet's spectrum.  We think that opacity due to TiO/VO is a natural explanation, but more work needs to be done to understand the abundances of these molecules at low pressure in these dynamic atmospheres, such as a non-equilibrium cloud condensation model \cp[see][and \S7.1]{Helling06}.  In green is a similar profile that utilizes a different method of treating the angular dependence of the incident flux \cp[see][]{Fortney07b}.  In blue is the day-side profile computed with TiO/VO opacity removed.

\mbox{Figure~\ref{209}}\emph{b} shows the spectra for these three models.  The flux levels at 4.5 and 5.8 $\mu$m are so large that they can only be caused by a temperature inversion.  This is similar to what was found at 8 $\mu$m for \hh\ by \citet{Harrington07}.  In the red model the temperature gradient at several millibars is not steep, so the emission features are quite weak.  Emission features are more prominent in the green model.  Both models with a temperature inversion also fit the 2.2 $\mu$m constraint from \citet{Rich03b}.

\citet{Burrows07c} present a nice fit to the \citet{Knutson08} observations.  With their models they explore two additional parameters not found in our models.  Most importantly, since they include a sink of energy on the day side, this cools their \hd\ atmosphere significantly (down to 1000 K) at the pressures responsible for emission in the 3.6$\mu$m band, leading to much less flux in this band.  This enables them to simultaneously fit the 3.6 and 4.5 $\mu$m band observations.  In addition, the opacity source that causes their temperature inversion has an optical depth for absorption that is similar to what they predict for TiO/VO, but it is not TiO/VO specifically.  Similar to our model, \citet{Burrows07c} have trouble simultaneously fitting the 5.8 $\mu$m and 8.0 $\mu$m observations.  Due to strong opacity due to water, emission at 24 $\mu$m emerges from high in the atmosphere (see \mbox{Figure~\ref{ptau}}).  In our models, this pressure is within the 2000+ K hot stratosphere, whereas \citet{Burrows07c} finds a cooler upper atmosphere ($\sim$1500 K) above their inversion.  In a future work we will explore \emph{P-T} profiles that extend to lower atmospheric pressure, to better understand how this may effect our derived profiles.  The most striking aspect of the \citet{Knutson08} observations is the relatively cool temperature needed to fit the 3.6 and 8.0 $\mu$m bands, relative to those at 4.5 and 5.8 $\mu$m.  Although our day side fits shows room for improvement, it is exciting that this planet falls into a class with \hh.  The details of absorption of incident flux by atmospheres partially depleted in TiO/VO due to condensation must be worked out in more detail before it is clear to what degree the TiO/VO abundances in the atmosphere of \hd\ are anomalously high.

\hd\ is a long-term target for the detection of optical flux from the \emph{MOST} satellite.  \ct{Rowe07} find a new 1$\sigma$ upper limit to the planet-to-star flux ratio.  \mbox{Figure~\ref{209}}\emph{b} shows that both models of \hd, with and without a temperature inversion, are consistent with this limit.  For the pL Class model, this is due to strong optical pressure-broadened lines of Na and K, as has been expected for some time \cp{Sudar00,SWS00}.  For the pM Class model TiO/VO opacity leads to less scattering of stellar flux, and TiO/VO band emission is not bright enough to lead to a detection.  It is a bit disappointing that \emph{MOST} bandpass ($\sim$400-680 nm) happens to coincide with a minimum in planetary flux.  However, the \emph{COROT} red bandpass and \emph{Kepler} band extend to redder wavelengths ($\sim1000$ nm and $\sim850$ nm respectively) such that the detection of planetary optical flux by these satellites for pM and pL Class planets should be achievable.

The opacity of TiO/VO leads to yet another explanation for the weaker than expected sodium absorption feature seen by \citet{Charb02}.  Although condensates are still an attractive option \cp{Fortney03,Fortney05c}, if TiO/VO are found on the day side of the planet's limb, opacity due to these molecules from $\sim$ 400 to 1000 nm could swamp the expected Na absorption feature at 589 nm.  If the opacity on the limb is everywhere larger in the optical this would lead to weaker, narrower Na absorption peak than expected by the first-generation models \cp{SS00,Brown01,Hubbard01}.  This effect can be seen in calculations of \ct{Barman02}, whose description of atmospheric chemistry at that time left addition absorbers of optical flux in the upper atmosphere.  It is not clear if TiO/VO would be found in abundance at the planet's limb, but this is an avenue worthy of further exploration.

\subsection{Other Planets}
We can examine the atmospheres of likely pM Class planets from \mbox{Figure~\ref{flux}}.  Since the OGLE planets will be difficult to detect in the infrared, we show \emph{P-T} profiles for WASP-1b \cp{Collier07,Charb07}, TrES-4b \cp{Mandushev07}, TrES-3b \cp{Odonovan07}, and $\upsilon$ And b \cp{Butler97,Fuhrmann98} in \mbox{Figure~\ref{pt3}}.  For these profiles the temperature of the adiabatic deep atmosphere at tens to hundred of bars (the top of the interior adiabat) has been chosen to be consistent with the measured planetary radius, assuming a Jupiter/Saturn like interior abundance of ice/rock \cp*[25 \me, see][]{Saumon04}.  The chosen adiabats do not rely on any specific evolution model---for a given planet mass and composition only one particular internal adiabat will match the planet's measured radius \cp[see, e.g.,][Figure 1]{Hubbard01}.  For the large-radii transiting planets, deep radiative zones extending down to $\sim$1 kbar are not consistent with the relatively warm interiors implied by these radii.  For $\upsilon$ And b, which has a mass similar to that of \hd, radius and surface gravity values for \hd\ were used.  The temperature of the upper atmosphere of these models are a function of the incident stellar flux, which decreases in magnitude from WASP-1b to $\upsilon$ And b.  We note that for these hotter pM Class planets, the day-side profiles are everywhere warmer than that required for Ti/V condensation, such that opacity due to TiO/VO should not be considered at all anomalous.  For comparison we also plot a pL Class model for potential transition object HAT-P-1b \cp{Bakos07a,Winn07b}, which transits one member of a main sequence G star binary system.  We highlight it here because the binary nature of the system makes it a good candidate for follow-up observations that rely on relative photometry.

We can examine the computed planet-to-star flux ratios for these models.  These are shown in \mbox{Figure~\ref{rat}}, using the same color scheme.  Most striking is TrES-3b, whose large planet-to-star radius ratio leads to large flux ratios reaching 0.01 at longer mid-IR wavelengths, as seen in \mbox{Figure~\ref{rat}}\emph{a}.  The pM Class models have fairly shallow temperature gradients in the hot stratosphere, leading to relatively week emission features in the spectra.  A comparison with HAT-P-1b shows that weak water emission features are seen in the pM Class planets at wavelengths where HAT-P-1b shows water absorption bands.  The feature at 4.5 $\mu$m is due predominantly to absorption in the stellar atmosphere, along with planetary emission due to CO.  The  incident stellar spectra are tailored for each planet's parent star and orbital distance, and are taken from \ct{Hauschildt99}.

\mbox{Figure~\ref{rat}}\emph{b} shows ratios in the optical and near infrared.  As has been pointed out previously, for these hot atmospheres thermal emission contributes significantly to the optical flux \cp{Fortney05,Seager05,Lopez07}.  These pM Class planets will be excellent targets for red-optical secondary eclipse observations, as the planet-to-star flux ratio reaches $\sim10^{-4}$.  It is important to note that this optical flux is \emph{thermal emission}, not scattered incident flux.  For the TrES-4b model (\emph{red}), the vertical bar indicates that by 0.45 $\mu$m, thermal flux is already \emph{10 times greater} than reflected light.  Thermal flux dominates at all longer optical wavelengths.  Even for pL Class model for HAT-P-1b, thermal emission is 10 times greater than reflected light at 0.73 $\mu$m.  If one wanted to observe true ``reflected light'' from these atmospheres, the blue spectral regions should be targeted.  In a future work we will examine geometric albedos and lightcurves for these planets to understand current and future data from spacecraft like \emph{MOST}, \emph{Kepler}, and \emph{COROT}.  \ct{Lopez07} have recently investigated prospects for the optical detection thermal emission from the ``very hot Jupiters'' in some detail and find the observations promising, even from the ground.  They calculate exposure times necessary for detection using blackbody planet models and also those of \ct{Hubeny03}.  They highlight OGLE-TR-56b and OGLE-TR-132b, the hottest planets yet detected, as the most promising candidates for secondary eclipse detections.

\section{Planetary Classification}
Previously \ct{Sudar00,Sudar03}  examined the
atmospheres and spectra of giant planets as a function of orbital
distance, and classified these atmospheres in terms of the
condensates that may be present. The pL Class that we suggest
here overlaps the Sudarsky et al. Class IV and V planets,
which are both dominated by alkali absorption. Class V
planets are distinguished by having silicate and iron clouds in the visible
atmosphere. However, it is not yet clear what effect these clouds
will have on hot Jupiter spectra and energy balance, because
particle sizes, cloud vertical distribution, and even composition
(MgSiO$_3$ vs. Mg$_2$SiO$_4$, which have very different optical properties)
are not known. Given that the derived \emph{P-T} profiles and
cloud condensation curves for pL Class atmospheres are somewhat
parallel (see Figure 2) there may be a relatively narrow
region of irradiation level (and \te) where clouds are present in
the visible atmosphere, rather than down at hundreds of bars.
The pM Class are planets with atmospheres hotter than those
considered by Sudarsky et al.

Although planetary classification based on dominant condensates
is certainly attractive, in practice this may by difficult to
implement. Clouds are often gray scatters over broad wavelength
ranges, so it may be challenging to recognize their presence and definitively identify
the composition of a given cloud deck if one is suspected, particularly with data extending 
over a limited spectral range. In
our own solar system the visible clouds of Jupiter and Saturn
are surely dominated by ammonia ice, as predicted by equilibrium
chemistry condensation, but there is as yet no spectroscopic
confirmation of this widely held belief.

A planetary classification scheme based on atomic and
molecular absorption features may be more practical to implement, and is in keeping with the tradition of basing classification directly on spectral features. This practice would be analogous to the spectral
MLT(Y) classification of cool dwarf stars and brown dwarfs.  The T
spectral type begins when methane is detected in the near infrared
\cp{Kirkpatrick05}, and it has been suggested that the arrival of ammonia absorption
in the near IR should delineate the arrival of the Y dwarfs \cp{Kirkpatrick05}.  Young, hot substellar objects and candidate planets directly imaged
at wider separations from their parent stars have and certainly will continue to be classified in the 
standard MLT(Y) scheme \cp[e.g.,][]{Chauvin05}.  Thus, extending
this classification scheme to all planets, near
and far from their parents stars, is a reasonable extension of the status quo.  The letters ``M'' and ``L'' are used to designate dwarfs at the
stellar/substellar boundary according to their spectral features, and for the planetary
objects with similar spectroscopic diagnostics as M and L dwarfs, we propose to use the
spectroscopic classification letters pM and pL.  Classes pT and pY would be natural extensions to cooler temperatures.

Planetary
classification will surely be complex, as some planets will be
dominated by absorbed stellar flux while others by their own
thermal emission, which will yield different \emph{P-T} profiles at a
given \te. The wavelength ranges at which planets will be most easily observed will also be 
a function of stellar glare and the sensitivity of future observatories.  Planets that are 
eccentric
will straddle two ore more classes, as suggested in Figure 1 for
HD 147506b and HD 17156b \cp[see also][for predictions at larger orbital separations]{Sudar05}.  All of these close-in planets will have, at a minimum of complexity, a day/night temperature difference that will make planetary classification difficult.  Obviously mass will be important as well
since with decreasing mass one eventually leaves the gas giant
regime.  \ct{Hansen07} have pointed out an interesting correlation of the known transiting planets with Safronov number; they appear to fall into two distinct classes.  This indicates that classification of planets surely will be complex and multi-dimensional.  Perhaps we will eventually find the \ct{Hansen07} classes possess different atmospheric abundances due to different accretion histories.  Thus we do not propose a detailed scheme, but rather suggest that the newly available transit data support a method of atmospheric classification based on molecular and atomic absorption, rather than the presence or absence of particular condensates.

\section{Discussion}
\subsection{TiO Chemistry Revisited}
Of course the boundary between the proposed pM and pL classes could
be somewhat indistinct, as the abundance of TiO/VO does not instantaneously
drop to zero upon Ti/V condensation. For TiO, \mbox{Figure~\ref{pt1}} shows that the mixing ratio falls by a factor of ten $\sim$40-180 K after initial Ti condensation at millibar to bar pressures (Lee \& Lodders, in prep). The spectra of early L dwarfs show a gradual weakening
the TiO and VO bands over progressing cooler sub-classes
\cp{Kirkpatrick99,Kirkpatrick05}; this points to transition
planets with depleted but still significant levels of TiO/VO
whose atmospheric temperatures are $\sim$100-200 K cooler than the
curve of Ti condensation at solar metallicity.  \hd\ is probably near the warm end of this transition region since \he\ shows no evidence for an inversion.  We note that at these same
temperatures, FeH and CrH are also prominent absorbers at
wavelengths blue-ward of 1.3 $\mu$m \cp{Kirkpatrick05}, before
being lost to Fe and Cr solid. Whether these molecules will
eventually help in pM and pL sub-typing will depend on the
quality of data available at these wavelengths. 

Since \hd\ appears to have a temperature inversion even at temperatures where the abundances of TiO is expected to be waning significantly, it is natural to inquire whether the abundance of TiO, may differ from the prediction from chemical equilibrium.  TiO is typically the major Ti-bearing gas at the \emph{P-T} conditions where Ti-bearing condensates are expected \cp[see][]{Lodders02b}. However, TiO$_2$ gas is next in abundance, and other Ti-bearing gases are also present.  With decreasing temperature (but before Ti-condensation), the TiO gas abundance decreases because abundances of other molecular Ti-bearing gases increase. The abundances of all Ti-bearing gases drop steeply when Ti-bearing condensates form.  However, if condensation is kinetically inhibited, the TiO gas abundances will drop much more gradually with decreasing temperatures, because the TiO is consumed only by other Ti-bearing gases, not condensates.

The abundance of TiO$_2$ gas is of interest because TiO$_2$ gas is a likely precursor of the TiO$_2$-building blocks in Ti-bearing condensates. The net reaction TiO + H$_2$O = TiO$_2$ + H$_2$ is independent of total pressure (there are the same number of molecules on both sides of the reaction). Under the relevant conditions here, the oxidation of the TiO radical is fast. For example, at 1750 K and 0.01 bar, oxidation of TiO to the equilibrium TiO$_2$ gas abundance takes less than a minute. This means that TiO abundances are unlikely to be influenced by any vertical mixing processes in planetary or brown dwarf atmospheres, and equilibrium TiO and TiO$_2$ abundances apply.

The formation of perovskite, CaTiO$_3$, the expected condensate at total pressures $< 0.02$ bar in a solar composition gas, requires reaction of TiO$_2$ with CaO. Preliminary calculations show that the oxidation of monatomic Ca to CaO gas proceeds on similar timescales as TiO to TiO$_2$ oxidation.  With respect to chemical changes by updraft mixing and quenching it must be emphasized that metal gas phase reactions are very fast so that equilibrium abundances established at higher temperature regimes will quickly re-adjust to the appropriate low temperature equilibria during upwelling.  Although the inhibition of the formation of Ti-bearing condensates should be kept in mind, another promising avenue for further work at this boundary may be further investigation into V-condensation.  As discussed in \S3.2, here we assume that VO is depleted with TiO, although strict chemical equilibrium would allow abundant VO at temperatures 200 K cooler than needed for initial Ti-condenation \ct{Lodders02b}.

At this time it would be unwise to rule out other absorbers such as non-equilibrium gases or condensates driven by photochemistry, which we do not consider here.  As \citet{Burrows07c} correctly point out, it may be possible for the abundance of a photochemically-derived absorber to scale as a function of irradiation level.  \citet{Liang04} have found that methane-derived hazes (which are present in the stratospheres of our solar system's giants) would not be stable for a variety of reasons, mainly because methane is not abundant at these high temperatures and low pressures, and haze particulate condensation is prevented.  \citet{Visscher06} also briefly investigated the photochemistry of C- and O-dominated species in \hd.  \ct{Marley07b} suggest that photochemistry involving multiple abundant elements, including sulfur, could be important for these planets.  Other authors \citep[e.~g.~][]{Yelle04} have investigated the photochemistry of the low density upper atmosphere of these planets as well.  Addition investigations in this area are surely needed.  

\subsection{Other Issues}
The models presented here are for day-side average planetary atmospheric structures and spectra.  For pL Class planets, redistribution of absorbed energy will likely lead to somewhat cooler atmospheres and lower fluxes than predicted here.  pM Class planetary atmospheres should be very close to a ``no redistribution model,'' that is hottest at the substellar point and becomes cooler toward the planetary limb.  A no redistribution day-side is more luminous than a uniform day-side average  \cp[see][who computed planet-wide average, day-side average, and no redistribution models of \hd\ and \T]{Barman05}.  Therefore our models are guides for understanding, and deviations from the predictions for specific planets are expected.  The pM Class planets could potentially be modestly brighter than we have shown.

To our knowledge, no models for the thermal evolution and contraction of pM Class planets have included TiO/VO opacity in their atmospheres.  Instead it has been assumed that Ti and V have condensed into clouds.  If it is found that WASP-1b, TrES-4b, and other pM Class planets have the atmospheric properties that we describe, this will mean that evolution models of these planets will need to be recomputed with new model atmosphere grids.  As shown in \S4.3, the atmospheric extension due to the hot stratospheric temperatures and strong TiO/VO opacity at optical wavelengths could lead to larger measured radii \cp[see also][]{Burrows07c}.

\section{Conclusions: Two Classes of Hot Atmospheres}
Though 1D radiative-convective equilibrium model atmospheres we have addressed the class of atmospheres, the pM Class, for which TiO and VO are extremely strong visible absorbers \cp{Hubeny03}.  This absorbed incident flux drives these planets to have hot ($\sim$2000+ K) stratospheres.  Therefore, these planets will appear very bright in the mid-infrared, with brightness temperatures larger than their equilibrium temperatures.  This is the case for \hh\ \cp{Harrington07}, and \hd\ \ct{Knutson08}.  In addition, these planets will have large day/night temperature contrasts because radiative time constants at photospheric pressures are much shorter than reasonable advective timescales.  The hottest point of the planet should be the substellar point, which absorbs the most flux, leading to perhaps negligible phase shift between the times of maximum measured thermal emission and when the day side is fully visible.  This appears to be the situation for planet $\upsilon$ And b, observed by \ct{Harrington06}.  Given that its irradiation level is intermediate between \hd\ and \hh\ we find that this planet is pM Class.  Due to the fast radiative times, the day side of these planets may have \emph{P-T} profiles that do not deviate much from radiative equilibrium models.  Although atmospheric dynamics will surely be vigorous, winds will be unable to advect gas before it cools to space.  For these planets, high irradiation, the presence of gaseous TiO and VO, a hot stratosphere, the location of the hottest atmosphere at the substellar point, and a large day/night temperature contrast all go hand-in-hand.

\begin{figure}
\epsscale{1.0}
\plotone{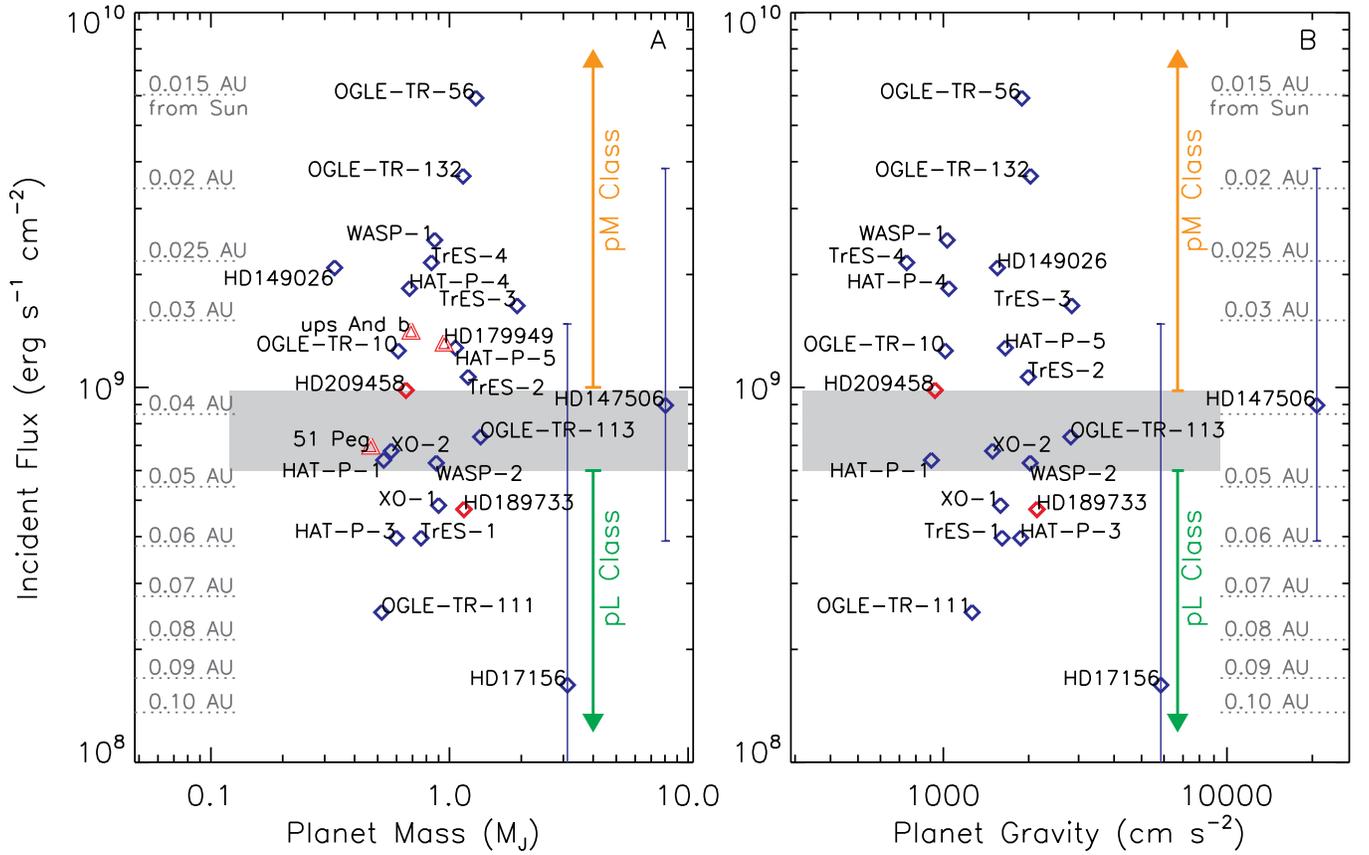}
\caption{Flux incident upon a collection of hot Jupiter planets.  At left is incident flux as a function of planet mass, and at right as a function of planet surface gravity.  In both figures the labeled dotted lines indicate the distance from the Sun that a planet would have to be to intercept this same flux.  Diamonds indicate the transiting planets while triangles indicate non-transiting systems  (with minimum masses plotted but unknown surface gravities).  Red color indicates that \emph{Spitzer} phase curve data are published, while blue color indicates there is no phase data.  The error bars for HD 147506 (HAT-P-2b) and HD 17156 indicate the variation in incident flux that the planets receive over their eccentric orbits.  Flux levels for pM Class and pL Class planets are shown, with the shaded region around $\sim$0.04-0.05 AU indicating the a possible transition region between the classes.  ``Hot Neptune'' GJ 436b experiences less intense insolation and is off the bottom of this plot at 3.2$\times$10$^7$ erg s$^{-1}$ cm$^{-2}$.
\label{flux}}
\end{figure}

\begin{figure}
\epsscale{1.0}
\plotone{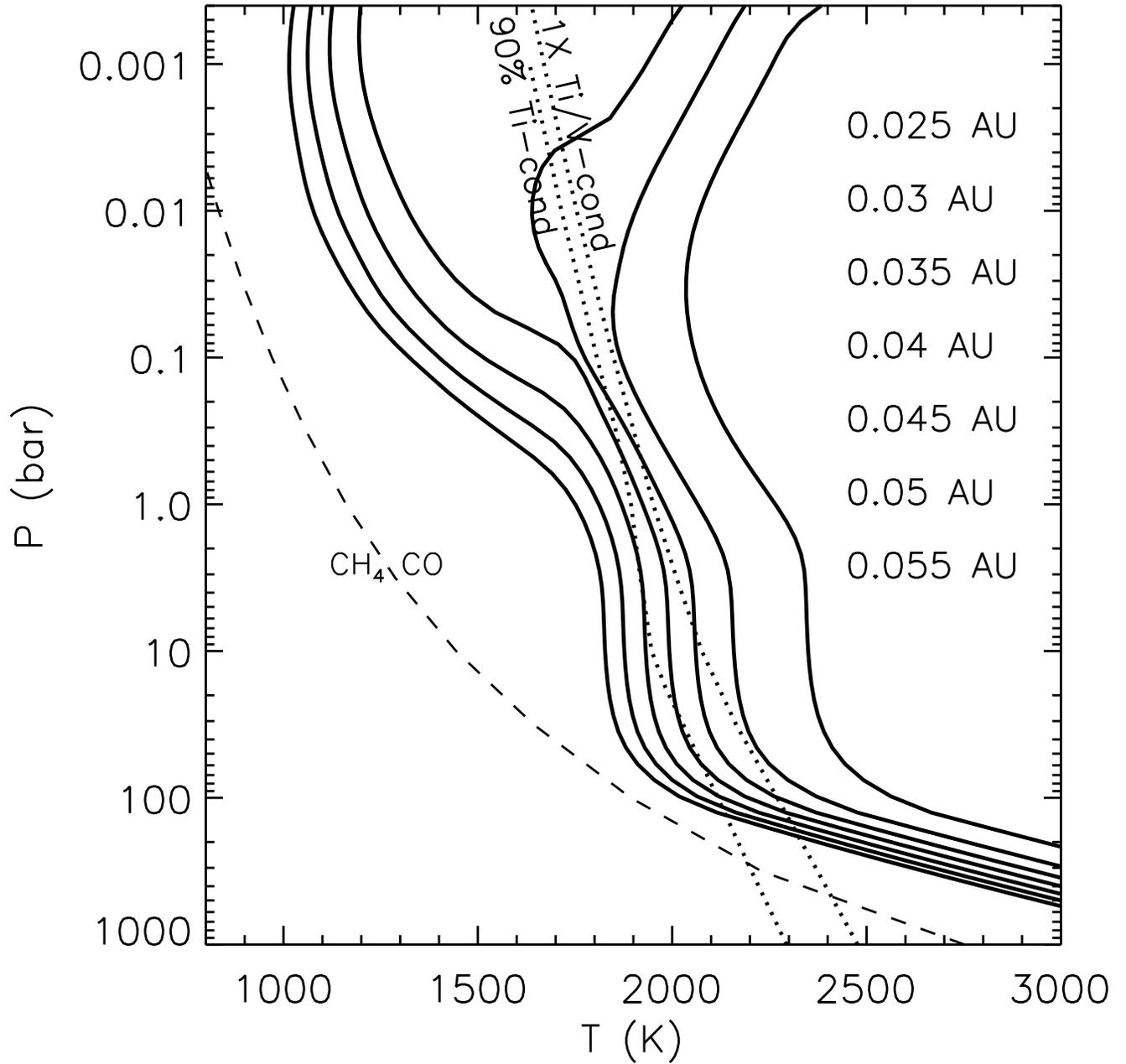}
\caption{Model \emph{P-T} profiles for planets with $g$=15 m s$^{-2}$ and $T_{\rm int}$=200 K at various distances (0.025 to 0.055 AU) from the Sun.  This $T_{\rm int}$ value is roughly consistent with that expected for a 1 \mj\ planet with a radius of 1.2 \rj.  Condensation curves are dotted lines and the curve where CO and CH$_4$ have equal abundances is dashed.  The 0.1X Ti-Cond curve shows where 90\% of the Ti has condensed out.  However, even then TiO is a major opacity source.  For none of these profiles has TiO/VO been artificially removed.  The kinks in the 0.035 AU profile are due to interpolation diffifulties, as the opacity drops significantly over a small temperature range.
\label{pt1}}
\end{figure}

\begin{figure}
\epsscale{1.0}
\plotone{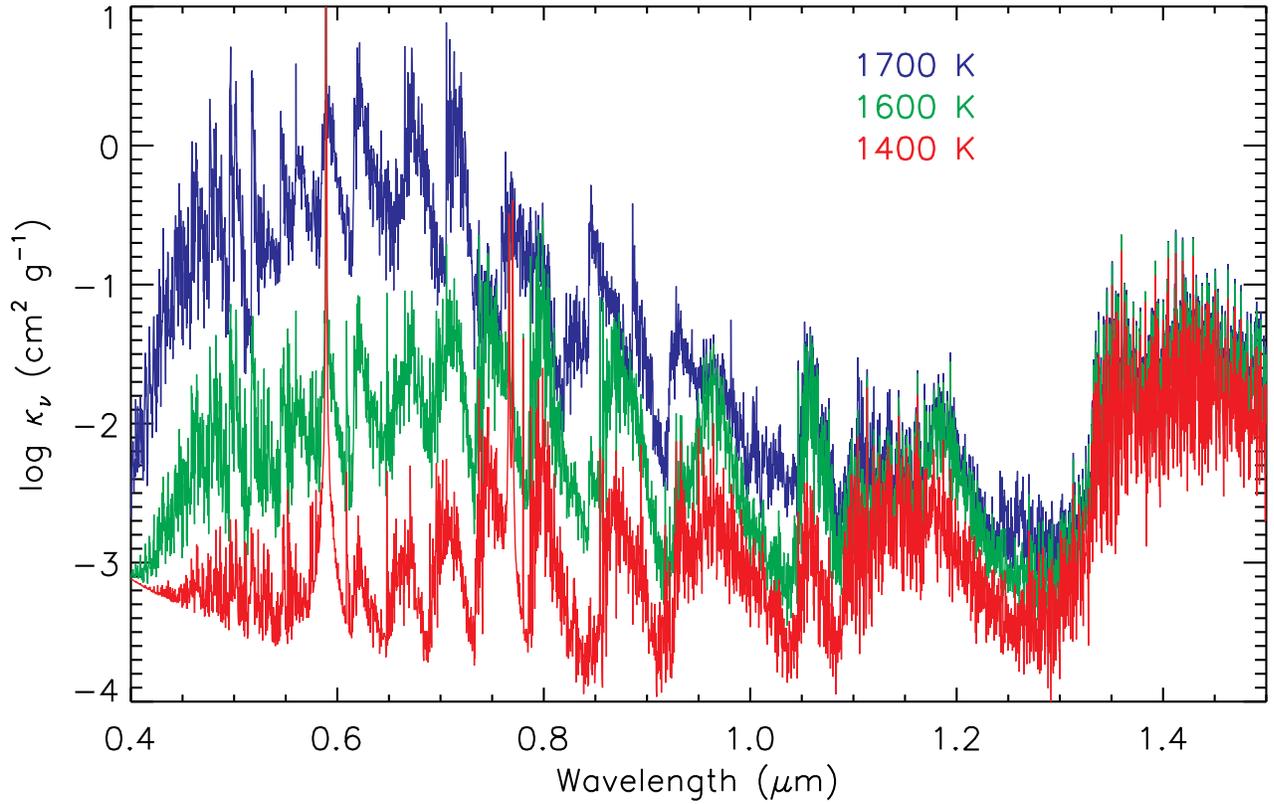}
\caption{Total abundance-weighted atmospheric opacity at $P$=1 mbar and $T$=1700 (blue), 1600 (green), 1400 (red) K.  The hottest temperature, 1700 K, is warmer than the temperature for Ti and V condensation, which begins to occur at 1670 K.  The intermediate temperature, 1600 K, is 33 K cooler than temperature at which Ti is 90\% condensed.  The optical opacity drops due to the removal of gaseous TiO and VO into Ti- and V-bearing solid condensates.  By 1400 K TiO and VO bands bands have given way to prominent water bands and alkali lines.  A strong water vapor band at 1.4 $\mu$m is readily seen at all three of these temperatures, and indicates the relatively modest change in absorption by water vapor with a 300 K drop in temperature.
\label{opac}}
\end{figure}

\begin{figure}
\epsscale{1.0}
\plotone{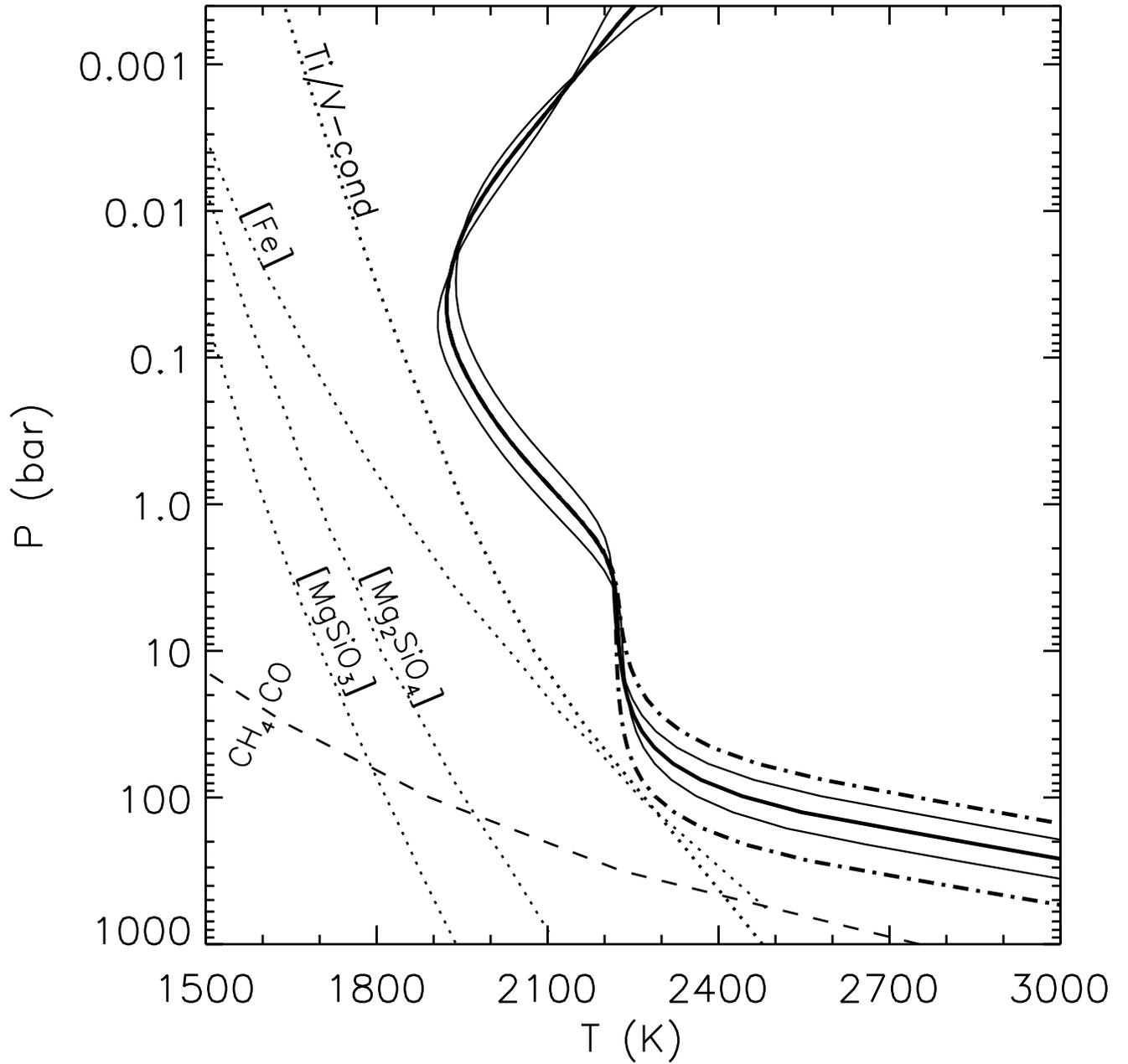}
\caption{Model \emph{P-T} profiles for planets at 0.028 AU from the Sun.  Effects of different values $T_{\rm int}$ and $g$ are shown.  Thick solid profile is for $g$=15 m s$^{-2}$ and $T_{\rm int}$=200 K.  Lower and higher dash-dot profiles are for $g$=15 m s$^{-2}$ with $T_{\rm int}$ values of 150 K and 250 K, respectively.  Lower and higher thin solid profiles are for $T_{\rm int}$=200 K with $g$ values of 25 m s$^{-2}$ and 9 m s$^{-2}$, respectively.
\label{pt2}}
\end{figure}

\begin{figure}
\epsscale{1.0}
\plotone{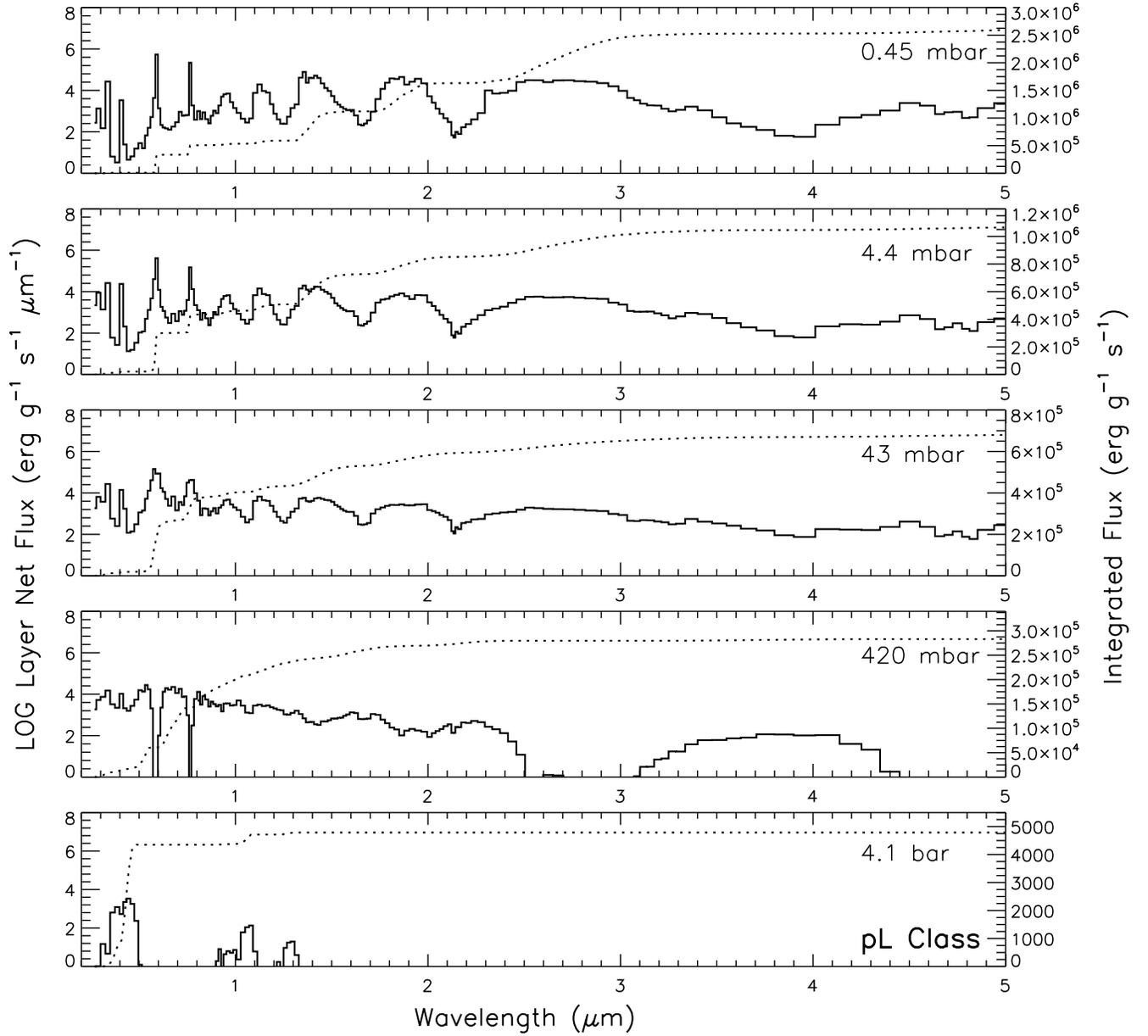}
\caption{Incident net ``flux'' (erg g$^{-1}$ s$^{-1}$ $\mu$m$^{-1}$, solid line, left ordinate) in five model layers for a cloud-free pL Class model with TiO/VO removed, $g$=15 m s$^{-2}$, at 0.05 AU from the Sun.The dotted line (right ordinate) illustrates the running integrated flux, evaluated from short to long wavelengths.  The layer \emph{integrated} flux is read at the intersection of the dotted line and the right ordinate.  Note the logarithmic scale on the left.  At 0.45 mbar, although the absorption due to neutral atomic alkalis are important, more flux is absorbed by water vapor.  Heating due to alkali absorption becomes relatively more important as pressure increases.  By 420 mbar, there is no flux left in the alkali line cores, and by 4.1 bar, nearly all the incident stellar flux has been absorbed. (After Marley \& Mckay, 1999.)
\label{v5}}
\end{figure}

\begin{figure}
\epsscale{1.0}
\plotone{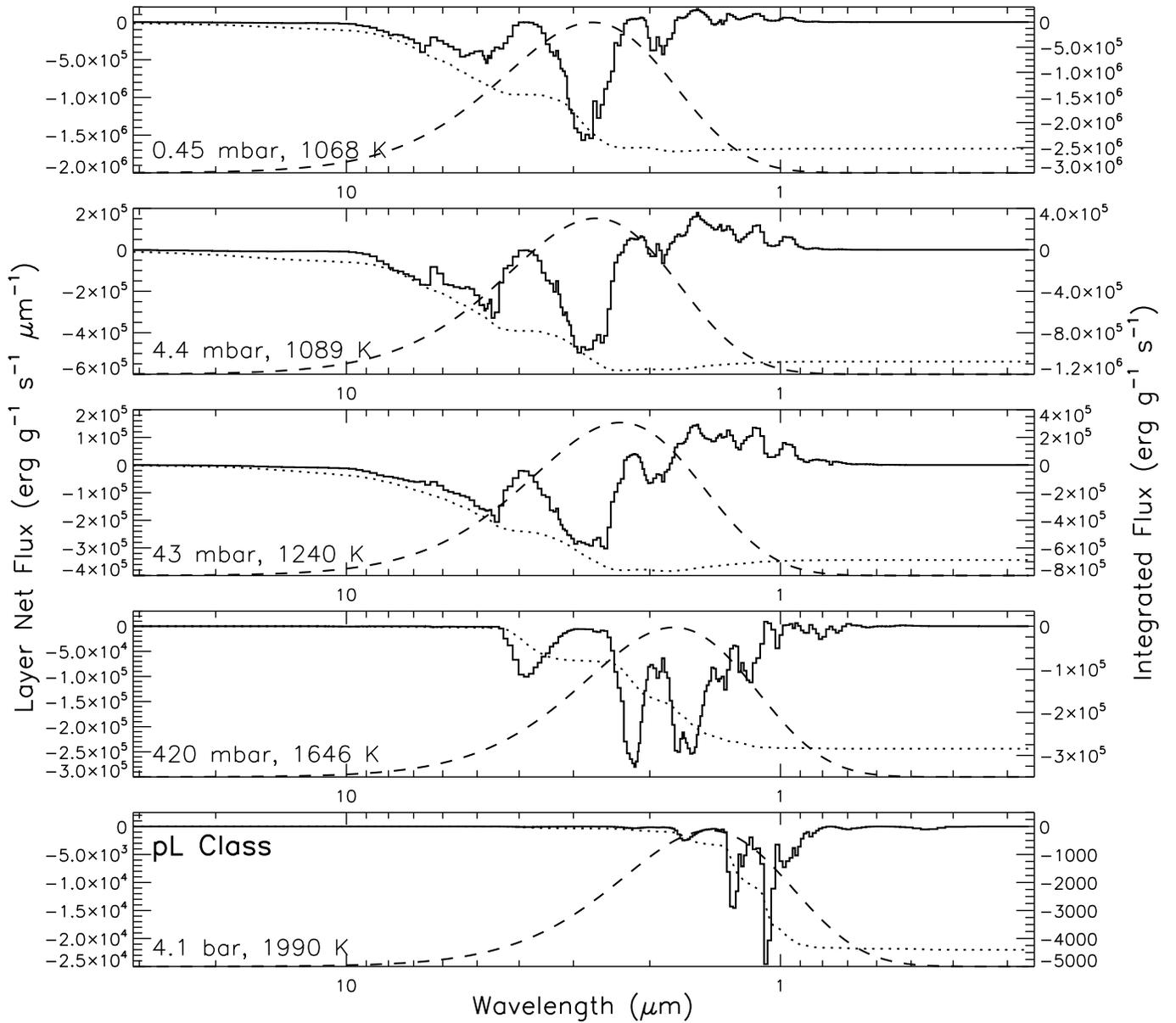}
\caption{Similar to \mbox{Figure~\ref{v5}}, but for thermal emission, from 30 to 0.26 $\mu$m, from long wavelengths to short.  Negative flux is emitted.  The dotted line is again integrated flux, evaluated from long to short wavelengths.  The negative value of a layer integrated flux in \mbox{Figure~\ref{v5}} equals the integrated flux here.  In addition, the scaled Planck function appropriate for each layer is shown as a dashed curve.  The temperature and pressure of each layer is labeled.  Cooling occurs mostly by way of water vapor, but also CO.  As the atmosphere cools with altitude progressively longer wavelength water bands dominate the layer thermal emission.
\label{i5}}
\end{figure}

\begin{figure}
\epsscale{1.0}
\plotone{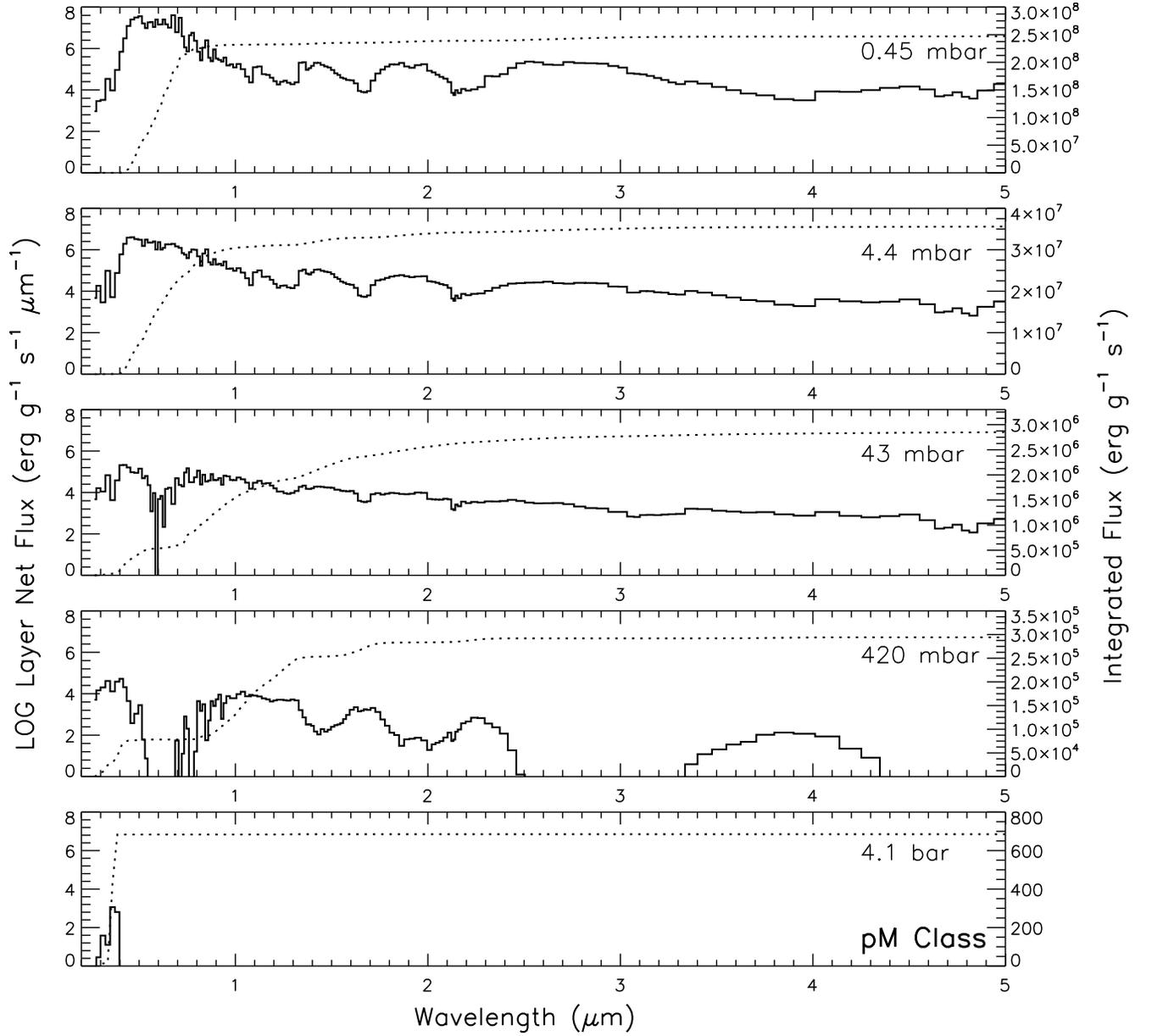}
\caption{Incident net ``flux'' (erg g$^{-1}$ s$^{-1}$ $\mu$m$^{-1}$, solid line, left ordinate) in five model layers for a pM Class model with $g$=15 m s$^{-2}$ at 0.03 AU from the Sun.  Only the left ordinate scale is the same as in \mbox{Figure~\ref{v5}}.  This pM Class model possesses a hot stratosphere.  At 0.45 and 4.4 mbar absorption of incident flux (heating) is due almost entirely to optical bands of TiO and VO.  Water vapor also absorbs flux as in the pL Class model, but its effect is comparatively swamped.  Also similar to the model shown in \mbox{Figure~\ref{v5}}, by 420 mbar most flux at optical wavelengths has already been absorbed.
\label{v3}}
\end{figure}

\begin{figure}
\epsscale{1.0}
\plotone{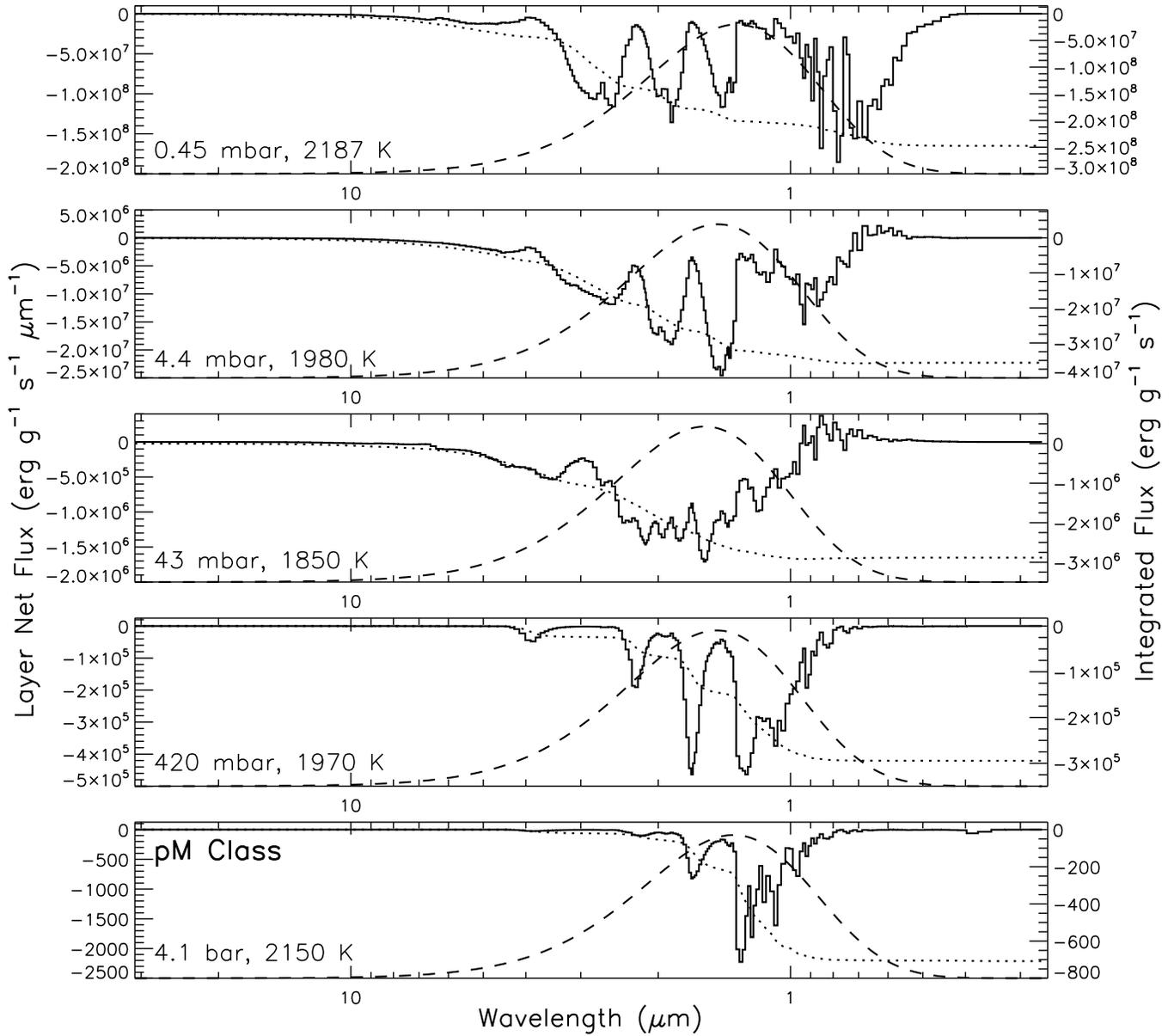}
\caption{Again, the model shown in \mbox{Figure~\ref{v3}}, but now for thermal emission.  At the topmost layer shown, the local Plank function must move to high temperatures to effectively radiate by means of bands of water, TiO, and VO.  At higher pressure layers, there is back-heating due to TiO and VO emission from above (seen especially at 4.4 and 43 mbar), along with cooling by water vapor.
\label{i3}}
\end{figure}

\begin{figure}
\epsscale{1.0}
\plotone{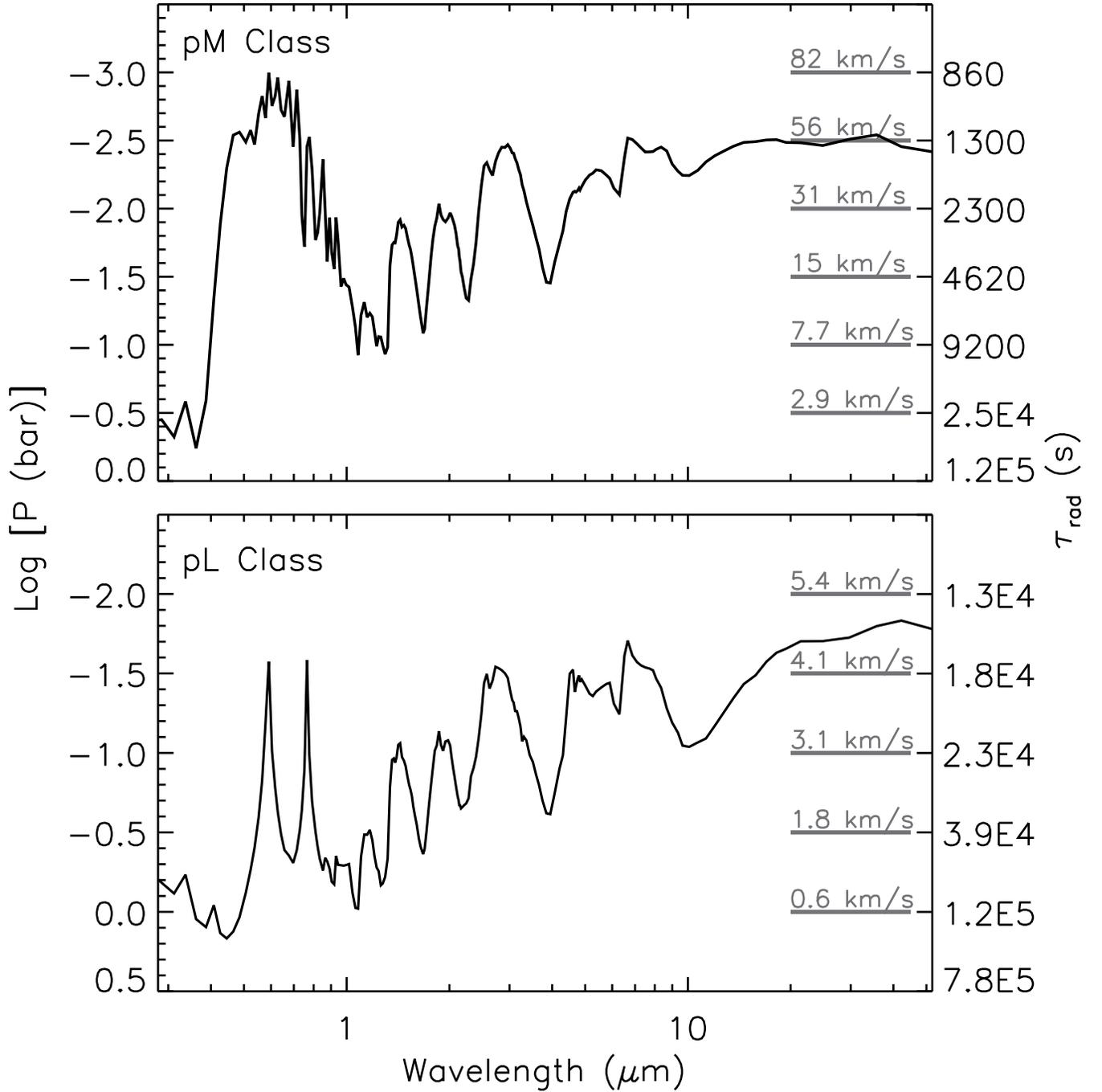}
\caption{Both figures show, as a function of wavelength, the atmospheric pressure that corresponds to a given brightness temperature.  Note the differences in the y-axes.  \emph{Top}: Planet at 0.03 AU which has a hot stratosphere induced by absorption by TiO/VO.  \emph{Bottom}: Planet at 0.05 AU that lacks a temperature inversion.  The right ordinate shows the corresponding radiative time constant at each major tick mark from the pressure axis.  Note that this right axis is not linear.  The labeled gray lines at right indicate an advective wind speed that would be necessary to give an advective time scale equal to the given radiative constant.  (See text.) For instance, in the top panel, at log $P$=-2.0, the radiative time constant is 2300 s, and an advection time of 2300 s would require a wind speed of 31 km s$^{-1}$ .
\label{ptau}}
\end{figure}

\begin{figure}
\epsscale{1.0}
\plotone{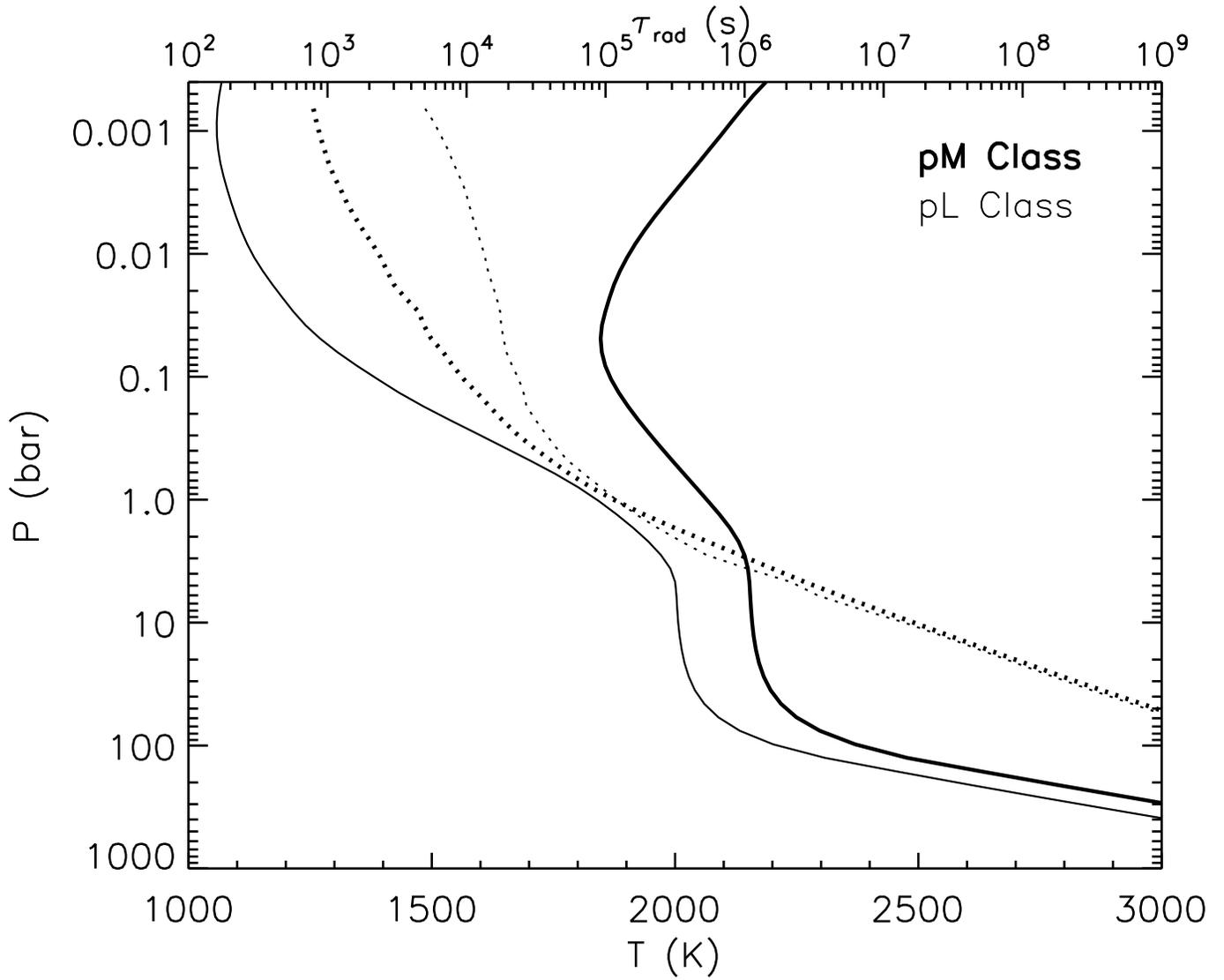}
\caption{Pressure-temperature (solid lines, bottom x-axis) and pressure-$\tau_{rad}$ profiles (dotted lines, top x-axis) for the models at 0.03 AU (pM Class, thick lines) and 0.05 AU (pL Class, thin lines, with TiO/VO removed).  TiO/VO has been removed when calculating the 0.05 AU profile.  Comparing this profile to the one at 0.05 AU in \mbox{Figure~\ref{pt1}}, where TiO/VO were not removed, shows slight differences around 1 bar, where equilibrium chemistry predicts a local increase in the TiO/VO abundances.
\label{PTtau}}
\end{figure}

\begin{figure}
\epsscale{1.0}
\plotone{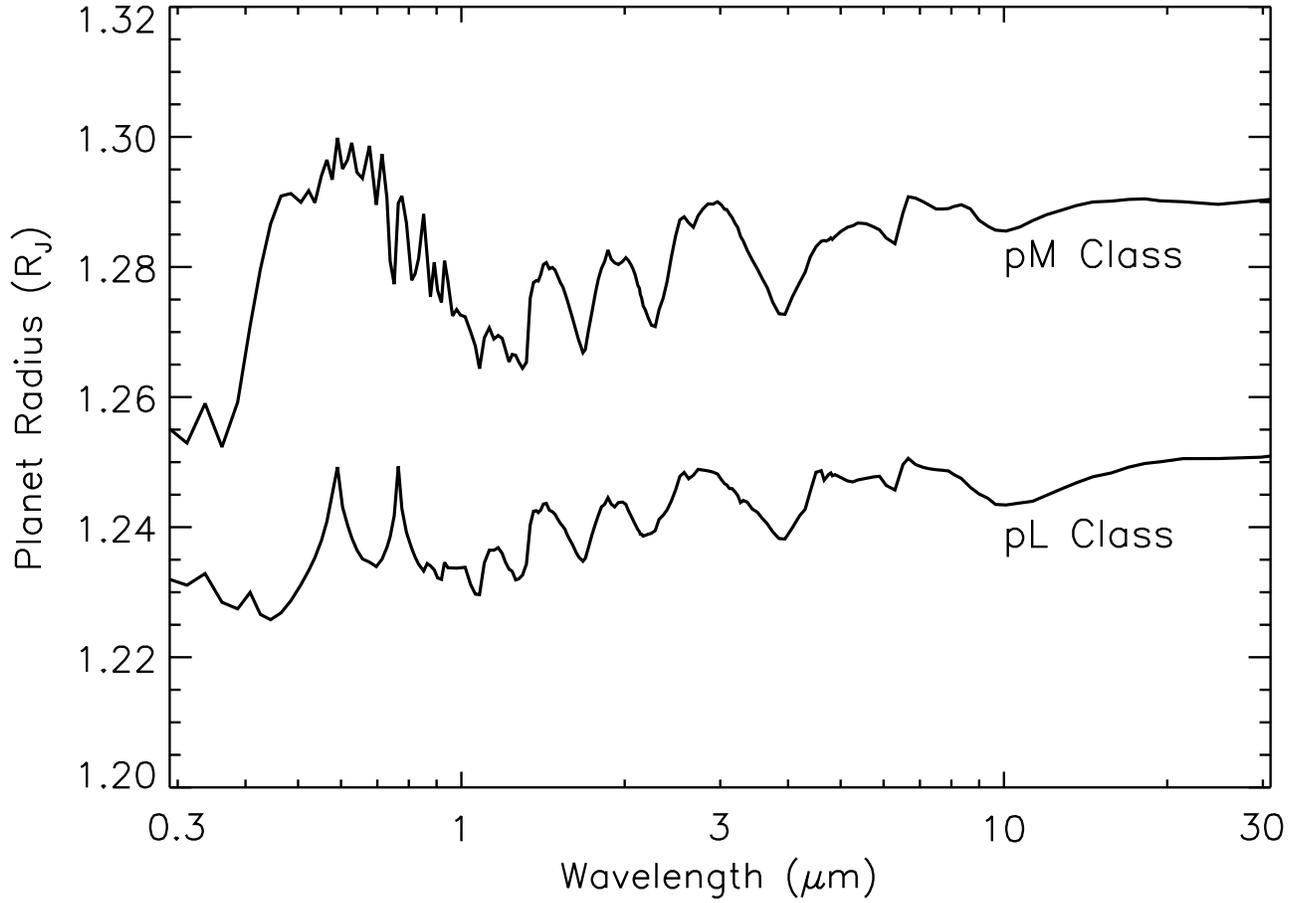}
\caption{Approximate radius one would observe as a function of wavelength for a pL Class and a pM Class planet with a 1 bar radius of 1.20 \rj\ and $g$=15 m s$^{-1}$. (See text for discussion.)
\label{rad}}
\end{figure}

\begin{figure}
\epsscale{1.0}
\plotone{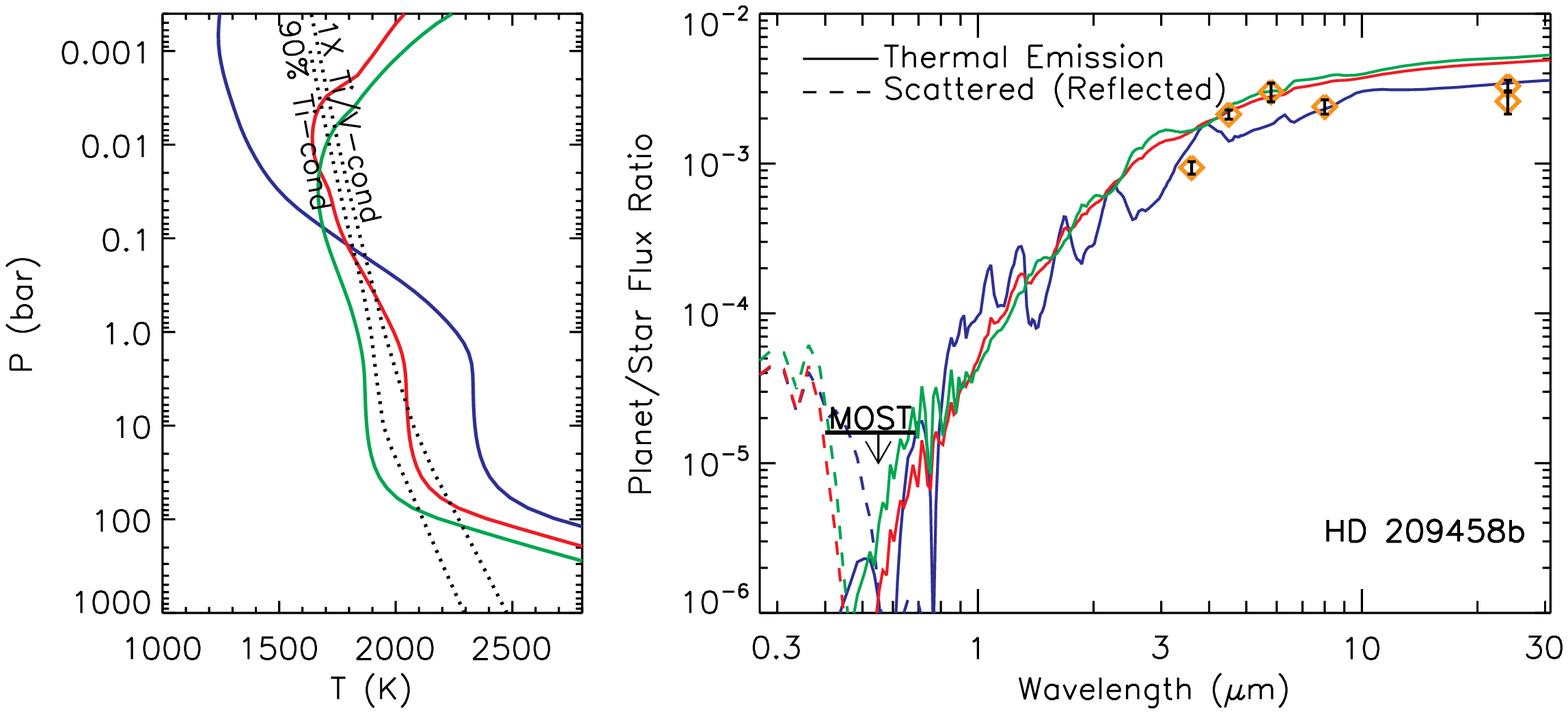}
\caption{\emph{Left}:  \hd\ atmospheric \emph{P-T} profiles for a model without TiO/VO opacity (blue) and two models with TiO/VO opacity (green and red).  The value of \ti\ is 200 K.  \emph{Right}:  Day-side low resolution model spectra for these profiles, along with observational data.  The \ct{Knutson08} data are shown from 3.6 to 8 $\mu$m.  At 24 $\mu$m the lower point is from \ct{Deming05b} and upper point is from D. Deming et al. (in prep).  Solid lines show thermal emission while dashed lines show scattered stellar flux.  The MOST 1$\sigma$ upper limit from $\sim$0.40-0.68 $\mu$m from \ct{Rowe07} is shown as horizontal black line.
\label{209}}
\end{figure}

\begin{figure}
\epsscale{1.0}
\plotone{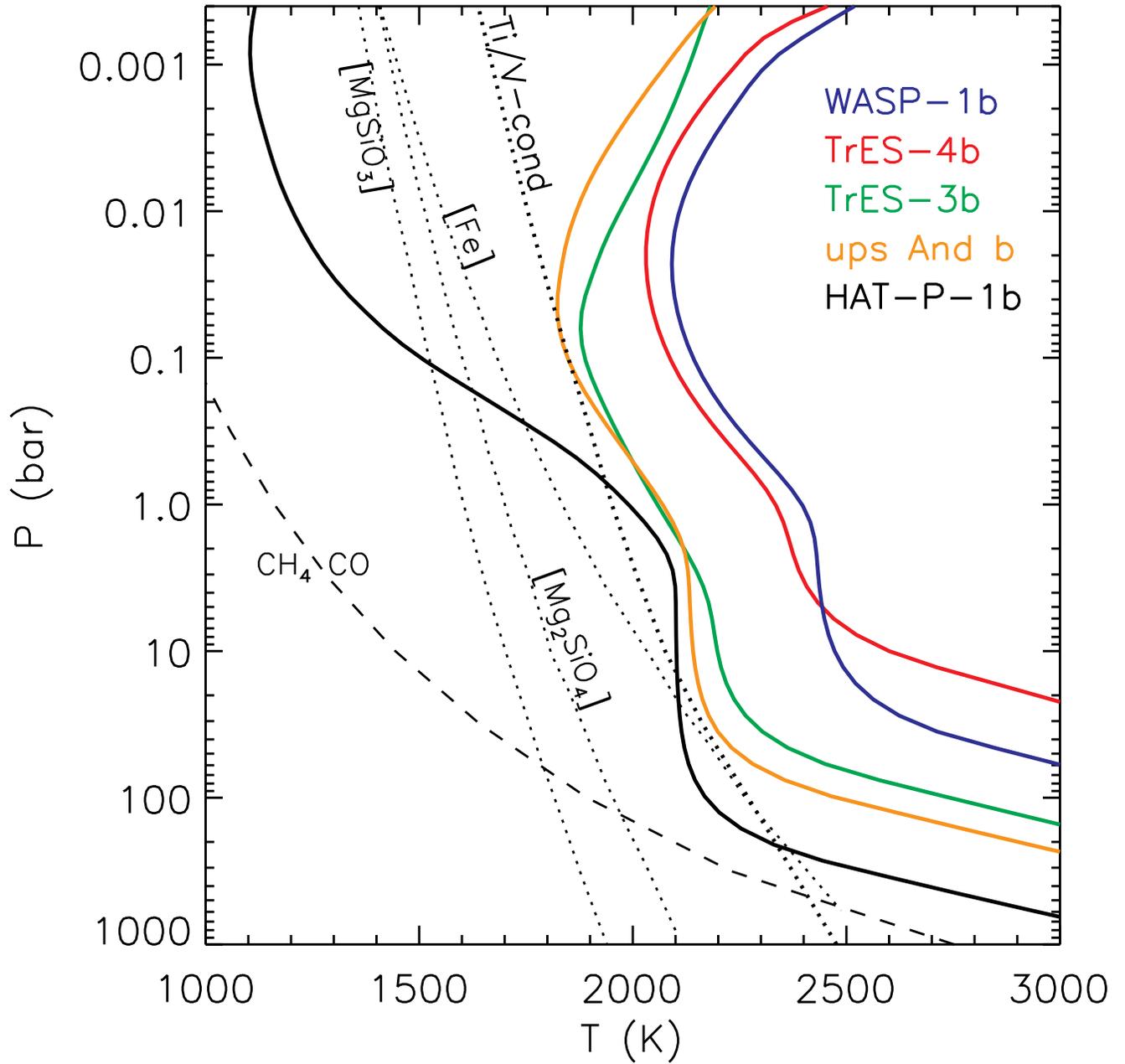}
\caption{Model \emph{P-T} profiles for five planets.  HAT-P-1b is likely pL class, which is the assumption here.  The other four planets are pM Class.
\label{pt3}}
\end{figure}

\begin{figure}
\epsscale{1.0}
\plotone{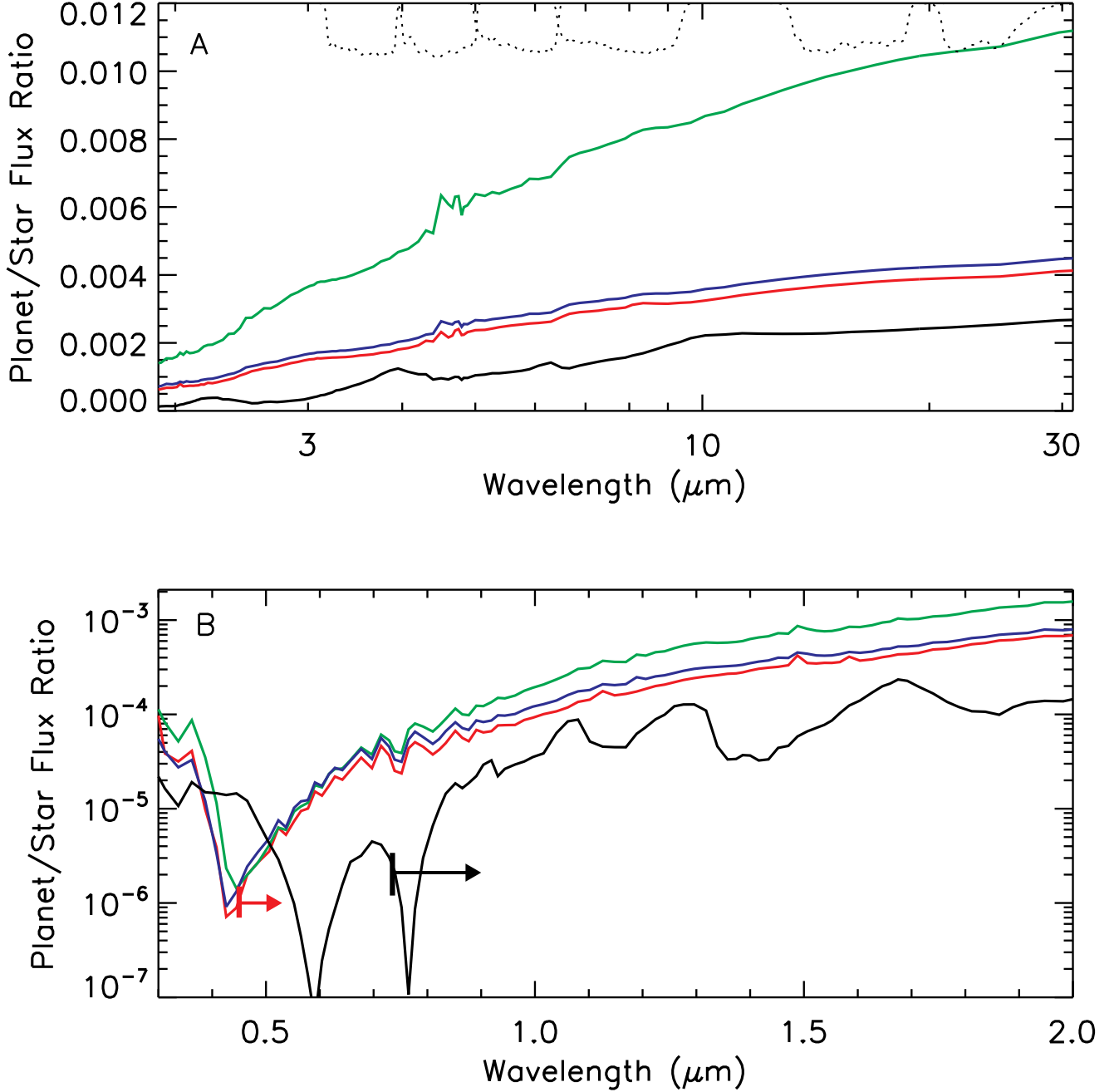}
\caption{Low resolution planet-to-star flux ratios for transiting planets shown in \mbox{Figure~\ref{pt3}}.  pM Class planets, due to their high incident fluxes are warmer and thus brighter.  Flux ratios may approach 10$^{-4}$ in the red optical.  Weak water emission bands are seen in the mid-infrared for the pM Class planets.  The bars with arrows indicate the wavelengths at which thermal emitted flux becomes 10 times greater than scattered incident flux, for the HAT-P-1b model (black) and the TrES-3b model (red).  pL Class fluxes red-ward of $\sim$0.7 $\mu$m are dominated by thermal emission, not reflected light.  pM Class fluxes red-ward of $\sim$0.4 $\mu$m  are dominated by thermal emission.
\label{rat}}
\end{figure}

On the other hand, we have shown that in the pL Class, dominated by absorption by H$_2$O, Na, and K, photospheric pressures and temperatures prevail such that advective timescales and radiative timescales are similar \cp[see also][]{Seager05}.  Since atmospheric dynamics will be important for the redistribution of energy, the consequences for the structure and thermal emission of these atmospheres will be quite complex.  The efficiency of energy redistribution will vary with planetary irradiation level, surface gravity, and rotation rate.  pL Class planets will have smaller day/night temperature contrasts and measurable phase shifts in thermal emission light curves that will be wavelength-dependent.  In addition, these planets may show variability in secondary eclipse depth \cp[e.g.~][]{Rauscher07}.  However, without a better understanding of the dynamics it is difficult to make detailed predictions at this time.  Secondary eclipse depths should range somewhere between values expected for a ``full redistribution'' model and inefficient redistribution.  The published secondary eclipse data for pL Class planets \T\ and \he\ are all consistent with this prediction \cp{Fortney05,Fortney07b}.  In addition, the 8 $\mu$m light curves for 51 Peg b and \hd\ \cp{Cowan07} and \he\ \cp{Knutson07b} are consistent with this prediction as well.  Examination of \mbox{Figure~\ref{flux}} shows that HD 179949b is a pM Class planet, and indeed \ct{Cowan07} found the largest phase variation it their small sample for this planet, but the unknown orbital inclination makes definitive conclusions difficult.

Additional observational results will soon help to test the models presented here.  We find that transiting planets WASP-1b, TrES-4b, TrES-3b, OGLE-Tr-10b, and TrES-2b will be in the pM Class, along with non-transiters $\upsilon$ And b and HD 179949.  The low-irradiation boundary of this class is not yet clear, and planet \hd\ shows that temperature inversions persist to irradiation levels where TiO/VO are expected to begin being lost to condensation \cp{Burrows07c}.  At still lower irradiation levels, the limited data for \he\ lead us to conclude it is pL Class \cp{Fortney07b}.  Secondary eclipse data for XO-2b, HAT-P-1b, and WASP-2b will be important is determining how temperature inverstions (and the TiO/VO abundances) wane with irradiation level.  Just as in dM and dL stars, the condensation of Ti and V is expected to be gradual process, so we fully expect transition objects between the distinct pM and pL class members.

We will soon have additional information that will help shed light on the atmosphere of \hd\.  The 8 $\mu$m light curve for \hd\ obtained by \ct{Cowan07} shows little phase variation.  However, if our theory connecting TiO/VO opacity and temperature inversions to large day/night contrasts is correct, we expect to see a large variation. Soon H.~Knutson and collaborators will obtain half-orbit light curves for \hd\ at 8 and 24 $\mu$m.  The quality should be comparable to that obtained by \citet{Knutson07b} for \he, and will put our theory to the test.  HD 147506, with an eccentricity of 0.517 \cp{Bakos07b}, and HD 17156, with an eccentricity of 0.67 \cp{Fischer07,Barbieri07}, will be extremely interesting cases as the flux they receive varies by factors of 9 and 26, respectively, between periapse and apoapse.  They should each spend part of their orbits as pL class and part as pM class.  This makes predictions difficult, but large day/night temperatures differences at apoapse are likely.

There has recently been considerable discussion on the relative merits of multi-dimensional dynamical models and 1D radiative-convective model atmospheres for these highly irradiated planets.  Both kinds of studies provide interesting predictions.  While it could be claimed that 1D radiative-convective models are unrealistic because they ``lack dynamics,'' they do include very detailed chemistry, vast opacity databases, and advanced non-gray radiative transfer, which all dynamics models for these planets lack.  Given the large diversity in predictions among the various dynamical models, which use a host of simplifications, the next step will be combining dynamics and radiative transfer, an idea which has been mentioned or advanced by a number of authors \cp{Seager05,Barman05,Fortney06b,Burrows06,Dobbs07}.  Eventually we will be able to give up 1D $f$-type parameters to treat the incident flux in a more realistic fashion for these exotic atmospheres.  We recently started working toward this goal in \ct[][see also \citealp{Dobbs07}]{Fortney06b}; work continues, and we believe it will have a promising future.  We think that the predictions we have made here with a 1D model will provide a framework for understanding the observations to come.  Additional observations for both pL and pM class planets, along with additional theoretical and modeling efforts, should further clarify our understanding.

\acknowledgements
We thank Adam Showman for valuable discussions and for help with the calculation of radiative time constants.   We thank Didier Saumon for providing the data files for Figure 3.  Jason Rowe kindly provided us with the MOST geometric albedo upper limit for \hd\ in advance of publication.  We also thank the referee and Jason Barnes for comments that improved the draft.  J.~J.~F.~acknowledges the support of a Spitzer Fellowship from NASA.  Work by K.~L.~is supported by NSF grant AST 0406963, and NASA grant NNG06GC26G.  Work by M.~S.~Marley is supported by the NASA Planetary Atmospheres Program.

%\bibliographystyle{apj}
%\bibliography{references}

\begin{thebibliography}{103}
\expandafter\ifx\csname natexlab\endcsname\relax\def\natexlab#1{#1}\fi

\bibitem[{{Ackerman} \& {Marley}(2001)}]{AM01}
{Ackerman}, A.~S. \& {Marley}, M.~S. 2001, \apj, 556, 872

\bibitem[{{Allard} {et~al.}(2001){Allard}, {Hauschildt}, {Alexander},
  {Tamanai}, \& {Schweitzer}}]{Allard01}
{Allard}, F., {Hauschildt}, P.~H., {Alexander}, D.~R., {Tamanai}, A., \&
  {Schweitzer}, A. 2001, \apj, 556, 357

\bibitem[{{Appleby} \& {Hogan}(1984)}]{Appleby84}
{Appleby}, J.~F. \& {Hogan}, J.~S. 1984, Icarus, 59, 336

\bibitem[{{Bakos} {et~al.}(2007{\natexlab{a}}){Bakos}, {Kov{\'a}cs}, {Torres},
  {Fischer}, {Latham}, {Noyes}, {Sasselov}, {Mazeh}, {Shporer}, {Butler},
  {Stefanik}, {Fern{\'a}ndez}, {Sozzetti}, {P{\'a}l}, {Johnson}, {Marcy},
  {Winn}, {Sip{\H o}cz}, {L{\'a}z{\'a}r}, {Papp}, \& {S{\'a}ri}}]{Bakos07b}
{Bakos}, G.~{\'A}., {Kov{\'a}cs}, G., {Torres}, G., {Fischer}, D.~A., {Latham},
  D.~W., {Noyes}, R.~W., {Sasselov}, D.~D., {Mazeh}, T., {Shporer}, A.,
  {Butler}, R.~P., {Stefanik}, R.~P., {Fern{\'a}ndez}, J.~M., {Sozzetti}, A.,
  {P{\'a}l}, A., {Johnson}, J., {Marcy}, G.~W., {Winn}, J.~N., {Sip{\H o}cz},
  B., {L{\'a}z{\'a}r}, J., {Papp}, I., \& {S{\'a}ri}, P. 2007{\natexlab{a}},
  \apj, 670, 826

\bibitem[{{Bakos} {et~al.}(2007{\natexlab{b}}){Bakos}, {Noyes}, {Kov{\'a}cs},
  {Latham}, {Sasselov}, {Torres}, {Fischer}, {Stefanik}, {Sato}, {Johnson},
  {P{\'a}l}, {Marcy}, {Butler}, {Esquerdo}, {Stanek}, {L{\'a}z{\'a}r}, {Papp},
  {S{\'a}ri}, \& {Sip{\H o}cz}}]{Bakos07a}
{Bakos}, G.~{\'A}., {Noyes}, R.~W., {Kov{\'a}cs}, G., {Latham}, D.~W.,
  {Sasselov}, D.~D., {Torres}, G., {Fischer}, D.~A., {Stefanik}, R.~P., {Sato},
  B., {Johnson}, J.~A., {P{\'a}l}, A., {Marcy}, G.~W., {Butler}, R.~P.,
  {Esquerdo}, G.~A., {Stanek}, K.~Z., {L{\'a}z{\'a}r}, J., {Papp}, I.,
  {S{\'a}ri}, P., \& {Sip{\H o}cz}, B. 2007{\natexlab{b}}, \apj, 656, 552

\bibitem[{{Baraffe} {et~al.}(2003){Baraffe}, {Chabrier}, {Barman}, {Allard}, \&
  {Hauschildt}}]{Baraffe03}
{Baraffe}, I., {Chabrier}, G., {Barman}, T.~S., {Allard}, F., \& {Hauschildt},
  P.~H. 2003, \aap, 402, 701

\bibitem[{{Barbieri} {et~al.}(2007){Barbieri}, {Alonso}, {Laughlin},
  {Almenara}, {Bissinger}, {Davies}, {Gasparri}, {Guido}, {Lopresti},
  {Manzini}, \& {Sostero}}]{Barbieri07}
{Barbieri}, M., {Alonso}, R., {Laughlin}, G., {Almenara}, J.~M., {Bissinger},
  R., {Davies}, D., {Gasparri}, D., {Guido}, E., {Lopresti}, C., {Manzini}, F.,
  \& {Sostero}, G. 2007, A\&A in press, ArXiv e-prints/0710.0898

\bibitem[{{Barman}(2007)}]{Barman07}
{Barman}, T. 2007, \apjl, 661, L191

\bibitem[{{Barman} {et~al.}(2005){Barman}, {Hauschildt}, \&
  {Allard}}]{Barman05}
{Barman}, T.~S., {Hauschildt}, P.~H., \& {Allard}, F. 2005, \apj, 632, 1132

\bibitem[{{Barman} {et~al.}(2002){Barman}, {Hauschildt}, {Schweitzer},
  {Stancil}, {Baron}, \& {Allard}}]{Barman02}
{Barman}, T.~S., {Hauschildt}, P.~H., {Schweitzer}, A., {Stancil}, P.~C.,
  {Baron}, E., \& {Allard}, F. 2002, \apjl, 569, L51

\bibitem[{{Bouchy} {et~al.}(2004){Bouchy}, {Pont}, {Santos}, {Melo}, {Mayor},
  {Queloz}, \& {Udry}}]{Bouchy04}
{Bouchy}, F., {Pont}, F., {Santos}, N.~C., {Melo}, C., {Mayor}, M., {Queloz},
  D., \& {Udry}, S. 2004, \aap, 421, L13

\bibitem[{{Bouchy} {et~al.}(2005){Bouchy}, {Udry}, {Mayor}, {Moutou}, {Pont},
  {Iribarne}, {da Silva}, {Ilovaisky}, {Queloz}, {Santos}, {S{\'e}gransan}, \&
  {Zucker}}]{Bouchy05}
{Bouchy}, F., {Udry}, S., {Mayor}, M., {Moutou}, C., {Pont}, F., {Iribarne},
  N., {da Silva}, R., {Ilovaisky}, S., {Queloz}, D., {Santos}, N.~C.,
  {S{\'e}gransan}, D., \& {Zucker}, S. 2005, \aap, 444, L15

\bibitem[{{Brown}(2001)}]{Brown01}
{Brown}, T.~M. 2001, \apj, 553, 1006

\bibitem[{{Burrows} {et~al.}(2007{\natexlab{a}}){Burrows}, {Hubeny}, {Budaj},
  \& {Hubbard}}]{Burrows07}
{Burrows}, A., {Hubeny}, I., {Budaj}, J., \& {Hubbard}, W.~B.
  2007{\natexlab{a}}, \apj, 661, 502

\bibitem[{{Burrows} {et~al.}(2007{\natexlab{b}}){Burrows}, {Hubeny}, {Budaj},
  {Knutson}, \& {Charbonneau}}]{Burrows07c}
{Burrows}, A., {Hubeny}, I., {Budaj}, J., {Knutson}, H.~A., \& {Charbonneau},
  D. 2007{\natexlab{b}}, \apjl, 668, L171

\bibitem[{{Burrows} {et~al.}(2004){Burrows}, {Hubeny}, {Hubbard}, {Sudarsky},
  \& {Fortney}}]{Burrows04}
{Burrows}, A., {Hubeny}, I., {Hubbard}, W.~B., {Sudarsky}, D., \& {Fortney},
  J.~J. 2004, \apjl, 610, L53

\bibitem[{{Burrows} {et~al.}(1997){Burrows}, {Marley}, {Hubbard}, {Lunine},
  {Guillot}, {Saumon}, {Freedman}, {Sudarsky}, \& {Sharp}}]{Burrows97}
{Burrows}, A., {Marley}, M., {Hubbard}, W.~B., {Lunine}, J.~I., {Guillot}, T.,
  {Saumon}, D., {Freedman}, R., {Sudarsky}, D., \& {Sharp}, C. 1997, \apj, 491,
  856

\bibitem[{{Burrows} {et~al.}(2000){Burrows}, {Marley}, \& {Sharp}}]{BMS}
{Burrows}, A., {Marley}, M.~S., \& {Sharp}, C.~M. 2000, \apj, 531, 438

\bibitem[{{Burrows} \& {Sharp}(1999)}]{Burrows99}
{Burrows}, A. \& {Sharp}, C.~M. 1999, \apj, 512, 843

\bibitem[{{Burrows} {et~al.}(2003){Burrows}, {Sudarsky}, \&
  {Hubbard}}]{Burrows03}
{Burrows}, A., {Sudarsky}, D., \& {Hubbard}, W.~B. 2003, \apj, 594, 545

\bibitem[{{Burrows} {et~al.}(2006){Burrows}, {Sudarsky}, \&
  {Hubeny}}]{Burrows06}
{Burrows}, A., {Sudarsky}, D., \& {Hubeny}, I. 2006, \apj, 650, 1140

\bibitem[{{Butler} {et~al.}(1997){Butler}, {Marcy}, {Williams}, {Hauser}, \&
  {Shirts}}]{Butler97}
{Butler}, R.~P., {Marcy}, G.~W., {Williams}, E., {Hauser}, H., \& {Shirts}, P.
  1997, \apjl, 474, L115+

\bibitem[{{Chabrier} \& {Baraffe}(2007)}]{Chabrier07c}
{Chabrier}, G. \& {Baraffe}, I. 2007, \apjl, 661, L81

\bibitem[{{Chamberlain} \& {Hunten}(1987)}]{ChambHunt}
{Chamberlain}, J.~W. \& {Hunten}, D.~M. 1987, Orlando FL Academic Press Inc
  International Geophysics Series, 36

\bibitem[{{Charbonneau} {et~al.}(2005){Charbonneau}, {Allen}, {Megeath},
  {Torres}, {Alonso}, {Brown}, {Gilliland}, {Latham}, {Mandushev}, {O'Donovan},
  \& {Sozzetti}}]{Charb05}
{Charbonneau}, D., {Allen}, L.~E., {Megeath}, S.~T., {Torres}, G., {Alonso},
  R., {Brown}, T.~M., {Gilliland}, R.~L., {Latham}, D.~W., {Mandushev}, G.,
  {O'Donovan}, F.~T., \& {Sozzetti}, A. 2005, ApJ, 626, 523

\bibitem[{{Charbonneau} {et~al.}(2002){Charbonneau}, {Brown}, {Noyes}, \&
  {Gilliland}}]{Charb02}
{Charbonneau}, D., {Brown}, T.~M., {Noyes}, R.~W., \& {Gilliland}, R.~L. 2002,
  \apj, 568, 377

\bibitem[{{Charbonneau} {et~al.}(2007){Charbonneau}, {Winn}, {Everett},
  {Latham}, {Holman}, {Esquerdo}, \& {O'Donovan}}]{Charb07}
{Charbonneau}, D., {Winn}, J.~N., {Everett}, M.~E., {Latham}, D.~W., {Holman},
  M.~J., {Esquerdo}, G.~A., \& {O'Donovan}, F.~T. 2007, \apj, 658, 1322

\bibitem[{{Chauvin} {et~al.}(2005){Chauvin}, {Lagrange}, {Dumas}, {Zuckerman},
  {Mouillet}, {Song}, {Beuzit}, \& {Lowrance}}]{Chauvin05}
{Chauvin}, G., {Lagrange}, A.-M., {Dumas}, C., {Zuckerman}, B., {Mouillet}, D.,
  {Song}, I., {Beuzit}, J.-L., \& {Lowrance}, P. 2005, A\&A, 438, L25

\bibitem[{{Collier Cameron} {et~al.}(2007){Collier Cameron}, {Bouchy},
  {H{\'e}brard}, {Maxted}, {Pollacco}, {Pont}, {Skillen}, {Smalley}, {Street},
  {West}, {Wilson}, {Aigrain}, {Christian}, {Clarkson}, {Enoch}, {Evans},
  {Fitzsimmons}, {Fleenor}, {Gillon}, {Haswell}, {Hebb}, {Hellier}, {Hodgkin},
  {Horne}, {Irwin}, {Kane}, {Keenan}, {Loeillet}, {Lister}, {Mayor}, {Moutou},
  {Norton}, {Osborne}, {Parley}, {Queloz}, {Ryans}, {Triaud}, {Udry}, \&
  {Wheatley}}]{Collier07}
{Collier Cameron}, A., {Bouchy}, F., {H{\'e}brard}, G., {Maxted}, P.,
  {Pollacco}, D., {Pont}, F., {Skillen}, I., {Smalley}, B., {Street}, R.~A.,
  {West}, R.~G., {Wilson}, D.~M., {Aigrain}, S., {Christian}, D.~J.,
  {Clarkson}, W.~I., {Enoch}, B., {Evans}, A., {Fitzsimmons}, A., {Fleenor},
  M., {Gillon}, M., {Haswell}, C.~A., {Hebb}, L., {Hellier}, C., {Hodgkin},
  S.~T., {Horne}, K., {Irwin}, J., {Kane}, S.~R., {Keenan}, F.~P., {Loeillet},
  B., {Lister}, T.~A., {Mayor}, M., {Moutou}, C., {Norton}, A.~J., {Osborne},
  J., {Parley}, N., {Queloz}, D., {Ryans}, R., {Triaud}, A.~H.~M.~J., {Udry},
  S., \& {Wheatley}, P.~J. 2007, \mnras, 375, 951

\bibitem[{{Cooper} \& {Showman}(2005)}]{CS05}
{Cooper}, C.~S. \& {Showman}, A.~P. 2005, \apjl, 629, L45

\bibitem[{{Cooper} \& {Showman}(2006)}]{CS06}
---. 2006, \apj, 649, 1048

\bibitem[{{Cowan} {et~al.}(2007){Cowan}, {Agol}, \& {Charbonneau}}]{Cowan07}
{Cowan}, N.~B., {Agol}, E., \& {Charbonneau}, D. 2007, \mnras, 379, 641

\bibitem[{{Cushing} {et~al.}(2007){Cushing}, {Marley}, {Saumon}, {Kelly},
  {Vacca}, {Rayner}, {Freedman}, {Lodders}, \& {Roellig}}]{Cushing08}
{Cushing}, M.~C., {Marley}, M.~S., {Saumon}, D., {Kelly}, B.~C., {Vacca},
  W.~D., {Rayner}, J.~T., {Freedman}, R.~S., {Lodders}, K., \& {Roellig}, T.~L.
  2007, ApJ in press, ArXiv e-prints/0711.0801

\bibitem[{{Deming} {et~al.}(2007){Deming}, {Harrington}, {Laughlin}, {Seager},
  {Navarro}, {Bowman}, \& {Horning}}]{Deming07}
{Deming}, D., {Harrington}, J., {Laughlin}, G., {Seager}, S., {Navarro}, S.~B.,
  {Bowman}, W.~C., \& {Horning}, K. 2007, \apjl, 667, L199

\bibitem[{{Deming} {et~al.}(2006){Deming}, {Harrington}, {Seager}, \&
  {Richardson}}]{Deming06}
{Deming}, D., {Harrington}, J., {Seager}, S., \& {Richardson}, L.~J. 2006,
  \apj, 644, 560

\bibitem[{{Deming} {et~al.}(2005){Deming}, {Seager}, {Richardson}, \&
  {Harrington}}]{Deming05b}
{Deming}, D., {Seager}, S., {Richardson}, L.~J., \& {Harrington}, J. 2005,
  Nature, 434, 740

\bibitem[{{Demory} {et~al.}(2007){Demory}, {Gillon}, {Barman}, {Bonfils},
  {Mayor}, {Mazeh}, {Queloz}, {Udry}, {Bouchy}, {Delfosse}, {Forveille},
  {Mallmann}, {Pepe}, \& {Perrier}}]{Demory07}
{Demory}, B.-O., {Gillon}, M., {Barman}, T., {Bonfils}, X., {Mayor}, M.,
  {Mazeh}, T., {Queloz}, D., {Udry}, S., {Bouchy}, F., {Delfosse}, X.,
  {Forveille}, T., {Mallmann}, F., {Pepe}, F., \& {Perrier}, C. 2007, \aap,
  475, 1125

\bibitem[{{Dobbs-Dixon} \& {Lin}(2007)}]{Dobbs07}
{Dobbs-Dixon}, I. \& {Lin}, D.~N.~C. 2007, ApJ in press, astro-ph/0704.3269

\bibitem[{{Fischer} {et~al.}(2007){Fischer}, {Vogt}, {Marcy}, {Butler}, {Sato},
  {Henry}, {Robinson}, {Laughlin}, {Ida}, {Toyota}, {Omiya}, {Driscoll},
  {Takeda}, {Wright}, \& {Johnson}}]{Fischer07}
{Fischer}, D.~A., {Vogt}, S.~S., {Marcy}, G.~W., {Butler}, R.~P., {Sato}, B.,
  {Henry}, G.~W., {Robinson}, S., {Laughlin}, G., {Ida}, S., {Toyota}, E.,
  {Omiya}, M., {Driscoll}, P., {Takeda}, G., {Wright}, J.~T., \& {Johnson},
  J.~A. 2007, \apj, 669, 1336

\bibitem[{{Fortney}(2005)}]{Fortney05c}
{Fortney}, J.~J. 2005, \mnras, 364, 649

\bibitem[{{Fortney} {et~al.}(2006{\natexlab{a}}){Fortney}, {Cooper}, {Showman},
  {Marley}, \& {Freedman}}]{Fortney06b}
{Fortney}, J.~J., {Cooper}, C.~S., {Showman}, A.~P., {Marley}, M.~S., \&
  {Freedman}, R.~S. 2006{\natexlab{a}}, \apj, 652, 746

\bibitem[{{Fortney} \& {Marley}(2007)}]{Fortney07b}
{Fortney}, J.~J. \& {Marley}, M.~S. 2007, \apjl, 666, L45

\bibitem[{{Fortney} {et~al.}(2005){Fortney}, {Marley}, {Lodders}, {Saumon}, \&
  {Freedman}}]{Fortney05}
{Fortney}, J.~J., {Marley}, M.~S., {Lodders}, K., {Saumon}, D., \& {Freedman},
  R. 2005, \apjl, 627, L69

\bibitem[{{Fortney} {et~al.}(2006{\natexlab{b}}){Fortney}, {Saumon}, {Marley},
  {Lodders}, \& {Freedman}}]{Fortney06}
{Fortney}, J.~J., {Saumon}, D., {Marley}, M.~S., {Lodders}, K., \& {Freedman},
  R.~S. 2006{\natexlab{b}}, \apj, 642, 495

\bibitem[{{Fortney} {et~al.}(2003){Fortney}, {Sudarsky}, {Hubeny}, {Cooper},
  {Hubbard}, {Burrows}, \& {Lunine}}]{Fortney03}
{Fortney}, J.~J., {Sudarsky}, D., {Hubeny}, I., {Cooper}, C.~S., {Hubbard},
  W.~B., {Burrows}, A., \& {Lunine}, J.~I. 2003, \apj, 589, 615

\bibitem[{{Freedman} {et~al.}(2007){Freedman}, {Marley}, \&
  {Lodders}}]{Freedman07}
{Freedman}, R.~S., {Marley}, M.~S., \& {Lodders}, K. 2007, ApJS in press,
  astro-ph/0706.2374

\bibitem[{{Fuhrmann} {et~al.}(1998){Fuhrmann}, {Pfeiffer}, \&
  {Bernkopf}}]{Fuhrmann98}
{Fuhrmann}, K., {Pfeiffer}, M.~J., \& {Bernkopf}, J. 1998, \aap, 336, 942

\bibitem[{{Goody} \& {Yung}(1989)}]{GoodyYung}
{Goody}, R.~M. \& {Yung}, Y.~L. 1989, {Atmospheric radiation : theoretical
  basis} (2nd ed., New York, NY: Oxford University Press, 1989)

\bibitem[{{Grillmair} {et~al.}(2007){Grillmair}, {Charbonneau}, {Burrows},
  {Armus}, {Stauffer}, {Meadows}, {Van Cleve}, \& {Levine}}]{Grillmair07}
{Grillmair}, C.~J., {Charbonneau}, D., {Burrows}, A., {Armus}, L., {Stauffer},
  J., {Meadows}, V., {Van Cleve}, J., \& {Levine}, D. 2007, \apjl, 658, L115

\bibitem[{{Hansen} \& {Barman}(2007)}]{Hansen07}
{Hansen}, B.~M.~S. \& {Barman}, T. 2007, ApJ in press, ArXiv e-prints/0706.3052

\bibitem[{{Harrington} {et~al.}(2006){Harrington}, {Hansen}, {Luszcz},
  {Seager}, {Deming}, {Menou}, {Cho}, \& {Richardson}}]{Harrington06}
{Harrington}, J., {Hansen}, B.~M., {Luszcz}, S.~H., {Seager}, S., {Deming}, D.,
  {Menou}, K., {Cho}, J.~Y.-K., \& {Richardson}, L.~J. 2006, Science, 314, 623

\bibitem[{{Harrington} {et~al.}(2007){Harrington}, {Luszcz}, {Seager},
  {Deming}, \& {Richardson}}]{Harrington07}
{Harrington}, J., {Luszcz}, S., {Seager}, S., {Deming}, D., \& {Richardson},
  L.~J. 2007, \nat, 447, 691

\bibitem[{{Hauschildt} {et~al.}(1999){Hauschildt}, {Allard}, {Ferguson},
  {Baron}, \& {Alexander}}]{Hauschildt99}
{Hauschildt}, P.~H., {Allard}, F., {Ferguson}, J., {Baron}, E., \& {Alexander},
  D.~R. 1999, \apj, 525, 871

\bibitem[{{Helling} \& {Woitke}(2006)}]{Helling06}
{Helling}, C. \& {Woitke}, P. 2006, \aap, 455, 325

\bibitem[{{Hubbard} {et~al.}(2001){Hubbard}, {Fortney}, {Lunine}, {Burrows},
  {Sudarsky}, \& {Pinto}}]{Hubbard01}
{Hubbard}, W.~B., {Fortney}, J.~J., {Lunine}, J.~I., {Burrows}, A., {Sudarsky},
  D., \& {Pinto}, P. 2001, \apj, 560, 413

\bibitem[{{Hubeny} {et~al.}(2003){Hubeny}, {Burrows}, \& {Sudarsky}}]{Hubeny03}
{Hubeny}, I., {Burrows}, A., \& {Sudarsky}, D. 2003, \apj, 594, 1011

\bibitem[{{Iro} {et~al.}(2005){Iro}, {Bezard}, \& {Guillot}}]{Iro05}
{Iro}, N., {Bezard}, B., \& {Guillot}, T. 2005, \aap, 436, 719

\bibitem[{{Kirkpatrick}(2005)}]{Kirkpatrick05}
{Kirkpatrick}, J.~D. 2005, \araa, 43, 195

\bibitem[{{Kirkpatrick} {et~al.}(1999){Kirkpatrick}, {Reid}, {Liebert},
  {Cutri}, {Nelson}, {Beichman}, {Dahn}, {Monet}, {Gizis}, \&
  {Skrutskie}}]{Kirkpatrick99}
{Kirkpatrick}, J.~D., {Reid}, I.~N., {Liebert}, J., {Cutri}, R.~M., {Nelson},
  B., {Beichman}, C.~A., {Dahn}, C.~C., {Monet}, D.~G., {Gizis}, J.~E., \&
  {Skrutskie}, M.~F. 1999, \apj, 519, 802

\bibitem[{{Knutson} {et~al.}(2007{\natexlab{a}}){Knutson}, {Charbonneau},
  {Allen}, {Burrows}, \& {Megeath}}]{Knutson08}
{Knutson}, H.~A., {Charbonneau}, D., {Allen}, L.~E., {Burrows}, A., \&
  {Megeath}, S.~T. 2007{\natexlab{a}}, ApJ in press, ArXiv e-prints/0709.3984

\bibitem[{{Knutson} {et~al.}(2007{\natexlab{b}}){Knutson}, {Charbonneau},
  {Allen}, {Fortney}, {Agol}, {Cowan}, {Showman}, {Cooper}, \&
  {Megeath}}]{Knutson07b}
{Knutson}, H.~A., {Charbonneau}, D., {Allen}, L.~E., {Fortney}, J.~J., {Agol},
  E., {Cowan}, N.~B., {Showman}, A.~P., {Cooper}, C.~S., \& {Megeath}, S.~T.
  2007{\natexlab{b}}, \nat, 447, 183

\bibitem[{{Konacki} {et~al.}(2003){Konacki}, {Torres}, {Jha}, \&
  {Sasselov}}]{Konacki03}
{Konacki}, M., {Torres}, G., {Jha}, S., \& {Sasselov}, D.~D. 2003, \nat, 421,
  507

\bibitem[{{Kornacki} \& {Fegley}(1986)}]{Kornacki86}
{Kornacki}, A.~S. \& {Fegley}, B.~J. 1986, Earth and Planetary Science Letters,
  79, 217

\bibitem[{{Langton} \& {Laughlin}(2007)}]{Langton07}
{Langton}, J. \& {Laughlin}, G. 2007, \apjl, 657, L113

\bibitem[{{Liang} {et~al.}(2004){Liang}, {Seager}, {Parkinson}, {Lee}, \&
  {Yung}}]{Liang04}
{Liang}, M., {Seager}, S., {Parkinson}, C.~D., {Lee}, A.~Y.-T., \& {Yung},
  Y.~L. 2004, \apjl, 605, L61

\bibitem[{{Lodders}(1999)}]{Lodders99}
{Lodders}, K. 1999, \apj, 519, 793

\bibitem[{{Lodders}(2002)}]{Lodders02b}
---. 2002, \apj, 577, 974

\bibitem[{{Lodders}(2003)}]{Lodders03}
---. 2003, \apj, 591, 1220

\bibitem[{{Lodders} \& {Fegley}(2002)}]{Lodders02}
{Lodders}, K. \& {Fegley}, B. 2002, Icarus, 155, 393

\bibitem[{{Lodders} \& {Fegley}(2006)}]{Lodders06}
---. 2006, {Astrophysics Update 2} (Springer Praxis Books, Berlin: Springer,
  2006)

\bibitem[{{L{\'o}pez-Morales} \& {Seager}(2007)}]{Lopez07}
{L{\'o}pez-Morales}, M. \& {Seager}, S. 2007, \apjl, 667, L191

\bibitem[{{Mandushev} {et~al.}(2007){Mandushev}, {O'Donovan}, {Charbonneau},
  {Torres}, {Latham}, {Bakos}, {Dunham}, {Sozzetti}, {Fern{\'a}ndez},
  {Esquerdo}, {Everett}, {Brown}, {Rabus}, {Belmonte}, \&
  {Hillenbrand}}]{Mandushev07}
{Mandushev}, G., {O'Donovan}, F.~T., {Charbonneau}, D., {Torres}, G., {Latham},
  D.~W., {Bakos}, G.~{\'A}., {Dunham}, E.~W., {Sozzetti}, A., {Fern{\'a}ndez},
  J.~M., {Esquerdo}, G.~A., {Everett}, M.~E., {Brown}, T.~M., {Rabus}, M.,
  {Belmonte}, J.~A., \& {Hillenbrand}, L.~A. 2007, \apjl, 667, L195

\bibitem[{{Marley} {et~al.}(2007){Marley}, {Fortney}, {Seager}, \&
  {Barman}}]{Marley07b}
{Marley}, M.~S., {Fortney}, J., {Seager}, S., \& {Barman}, T. 2007, in
  Protostars and Planets V, ed. B.~{Reipurth}, D.~{Jewitt}, \& K.~{Keil},
  733--747

\bibitem[{{Marley} \& {McKay}(1999)}]{MM99}
{Marley}, M.~S. \& {McKay}, C.~P. 1999, Icarus, 138, 268

\bibitem[{{Marley} {et~al.}(1996){Marley}, {Saumon}, {Guillot}, {Freedman},
  {Hubbard}, {Burrows}, \& {Lunine}}]{Marley96}
{Marley}, M.~S., {Saumon}, D., {Guillot}, T., {Freedman}, R.~S., {Hubbard},
  W.~B., {Burrows}, A., \& {Lunine}, J.~I. 1996, Science, 272, 1919

\bibitem[{{Marley} {et~al.}(2002){Marley}, {Seager}, {Saumon}, {Lodders},
  {Ackerman}, {Freedman}, \& {Fan}}]{Marley02}
{Marley}, M.~S., {Seager}, S., {Saumon}, D., {Lodders}, K., {Ackerman}, A.~S.,
  {Freedman}, R.~S., \& {Fan}, X. 2002, \apj, 568, 335

\bibitem[{{McKay} {et~al.}(1989){McKay}, {Pollack}, \& {Courtin}}]{Mckay89}
{McKay}, C.~P., {Pollack}, J.~B., \& {Courtin}, R. 1989, Icarus, 80, 23

\bibitem[{{O'Donovan} {et~al.}(2007){O'Donovan}, {Charbonneau}, {Bakos},
  {Mandushev}, {Dunham}, {Brown}, {Latham}, {Torres}, {Sozzetti}, {Kov{\'a}cs},
  {Everett}, {Baliber}, {Hidas}, {Esquerdo}, {Rabus}, {Deeg}, {Belmonte},
  {Hillenbrand}, \& {Stefanik}}]{Odonovan07}
{O'Donovan}, F.~T., {Charbonneau}, D., {Bakos}, G.~{\'A}., {Mandushev}, G.,
  {Dunham}, E.~W., {Brown}, T.~M., {Latham}, D.~W., {Torres}, G., {Sozzetti},
  A., {Kov{\'a}cs}, G., {Everett}, M.~E., {Baliber}, N., {Hidas}, M.~G.,
  {Esquerdo}, G.~A., {Rabus}, M., {Deeg}, H.~J., {Belmonte}, J.~A.,
  {Hillenbrand}, L.~A., \& {Stefanik}, R.~P. 2007, \apjl, 663, L37

\bibitem[{{Rauscher} {et~al.}(2007{\natexlab{a}}){Rauscher}, {Menou}, {Cho},
  {Seager}, \& {Hansen}}]{Rauscher07}
{Rauscher}, E., {Menou}, K., {Cho}, J.~Y.-K., {Seager}, S., \& {Hansen},
  B.~M.~S. 2007{\natexlab{a}}, \apjl, 662, L115

\bibitem[{{Rauscher} {et~al.}(2007{\natexlab{b}}){Rauscher}, {Menou}, {Seager},
  {Deming}, {Cho}, \& {Hansen}}]{Rauscher07b}
{Rauscher}, E., {Menou}, K., {Seager}, S., {Deming}, D., {Cho}, J.~Y.-K., \&
  {Hansen}, B.~M.~S. 2007{\natexlab{b}}, \apj, 664, 1199

\bibitem[{{Richardson} {et~al.}(2007){Richardson}, {Deming}, {Horning},
  {Seager}, \& {Harrington}}]{Richardson07}
{Richardson}, L.~J., {Deming}, D., {Horning}, K., {Seager}, S., \&
  {Harrington}, J. 2007, \nat, 445, 892

\bibitem[{{Richardson} {et~al.}(2003){Richardson}, {Deming}, \&
  {Seager}}]{Rich03b}
{Richardson}, L.~J., {Deming}, D., \& {Seager}, S. 2003, \apj, 597, 581

\bibitem[{{Rowe} {et~al.}(2007){Rowe}, {Matthews}, {Seager}, {Miller-Ricci},
  {Sasselov}, {Kuschnig}, {Guenther}, {Moffat}, {Rucinski}, {Walker}, \&
  {Weiss}}]{Rowe07}
{Rowe}, J.~F., {Matthews}, J.~M., {Seager}, S., {Miller-Ricci}, E., {Sasselov},
  D., {Kuschnig}, R., {Guenther}, D.~B., {Moffat}, A.~F.~J., {Rucinski}, S.~M.,
  {Walker}, G.~A.~H., \& {Weiss}, W.~W. 2007, ApJ submitted, ArXiv
  e-prints/0711.4111

\bibitem[{{Salby}(1996)}]{Salby}
{Salby}, M.~L. 1996, San Diego CA Academic Press Inc International Geophysics
  Series, 61

\bibitem[{{Saumon} \& {Guillot}(2004)}]{Saumon04}
{Saumon}, D. \& {Guillot}, T. 2004, \apj, 609, 1170

\bibitem[{{Saumon} {et~al.}(2006){Saumon}, {Marley}, {Cushing}, {Leggett},
  {Roellig}, {Lodders}, \& {Freedman}}]{Saumon06}
{Saumon}, D., {Marley}, M.~S., {Cushing}, M.~C., {Leggett}, S.~K., {Roellig},
  T.~L., {Lodders}, K., \& {Freedman}, R.~S. 2006, \apj, 647, 552

\bibitem[{{Saumon} {et~al.}(2007){Saumon}, {Marley}, {Leggett}, {Geballe},
  {Stephens}, {Golimowski}, {Cushing}, {Fan}, {Rayner}, {Lodders}, \&
  {Freedman}}]{Saumon07}
{Saumon}, D., {Marley}, M.~S., {Leggett}, S.~K., {Geballe}, T.~R., {Stephens},
  D., {Golimowski}, D.~A., {Cushing}, M.~C., {Fan}, X., {Rayner}, J.~T.,
  {Lodders}, K., \& {Freedman}, R.~S. 2007, \apj, 656, 1136

\bibitem[{{Seager} {et~al.}(2005){Seager}, {Richardson}, {Hansen}, {Menou},
  {Cho}, \& {Deming}}]{Seager05}
{Seager}, S., {Richardson}, L.~J., {Hansen}, B.~M.~S., {Menou}, K., {Cho},
  J.~Y.-K., \& {Deming}, D. 2005, \apj, 632, 1122

\bibitem[{{Seager} \& {Sasselov}(2000)}]{SS00}
{Seager}, S. \& {Sasselov}, D.~D. 2000, \apj, 537, 916

\bibitem[{{Seager} {et~al.}(2000){Seager}, {Whitney}, \& {Sasselov}}]{SWS00}
{Seager}, S., {Whitney}, B.~A., \& {Sasselov}, D.~D. 2000, \apj, 540, 504

\bibitem[{{Sharp} \& {Burrows}(2007)}]{Sharp07}
{Sharp}, C.~M. \& {Burrows}, A. 2007, \apjs, 168, 140

\bibitem[{{Showman} \& {Guillot}(2002)}]{Showman02}
{Showman}, A.~P. \& {Guillot}, T. 2002, \aap, 385, 166

\bibitem[{{Stevenson}(1985)}]{Stevenson85}
{Stevenson}, D.~J. 1985, Icarus, 62, 4

\bibitem[{{Sudarsky} {et~al.}(2003){Sudarsky}, {Burrows}, \&
  {Hubeny}}]{Sudar03}
{Sudarsky}, D., {Burrows}, A., \& {Hubeny}, I. 2003, \apj, 588, 1121

\bibitem[{{Sudarsky} {et~al.}(2005){Sudarsky}, {Burrows}, {Hubeny}, \&
  {Li}}]{Sudar05}
{Sudarsky}, D., {Burrows}, A., {Hubeny}, I., \& {Li}, A. 2005, \apj, 627, 520

\bibitem[{{Sudarsky} {et~al.}(2000){Sudarsky}, {Burrows}, \& {Pinto}}]{Sudar00}
{Sudarsky}, D., {Burrows}, A., \& {Pinto}, P. 2000, \apj, 538, 885

\bibitem[{{Swain} {et~al.}(2007){Swain}, {Bouwman}, {Akeson}, {Lawler}, \&
  {Beichman}}]{Swain07}
{Swain}, M.~R., {Bouwman}, J., {Akeson}, R., {Lawler}, S., \& {Beichman}, C.
  2007, ApJ in press, astro-ph/0702593

\bibitem[{{Tinetti} {et~al.}(2007){Tinetti}, {Vidal-Madjar}, {Liang},
  {Beaulieu}, {Yung}, {Carey}, {Barber}, {Tennyson}, {Ribas}, {Allard},
  {Ballester}, {Sing}, \& {Selsis}}]{Tinetti07}
{Tinetti}, G., {Vidal-Madjar}, A., {Liang}, M.-C., {Beaulieu}, J.-P., {Yung},
  Y., {Carey}, S., {Barber}, R.~J., {Tennyson}, J., {Ribas}, I., {Allard}, N.,
  {Ballester}, G.~E., {Sing}, D.~K., \& {Selsis}, F. 2007, \nat, 448, 169

\bibitem[{{Toon} {et~al.}(1989){Toon}, {McKay}, {Ackerman}, \&
  {Santhanam}}]{Toon89}
{Toon}, O.~B., {McKay}, C.~P., {Ackerman}, T.~P., \& {Santhanam}, K. 1989,
  Journal of Geophysical Research, 94, 16287

\bibitem[{{Visscher} {et~al.}(2006){Visscher}, {Lodders}, \&
  {Fegley}}]{Visscher06}
{Visscher}, C., {Lodders}, K., \& {Fegley}, B.~J. 2006, \apj, 648, 1181

\bibitem[{{Williams} {et~al.}(2006){Williams}, {Charbonneau}, {Cooper},
  {Showman}, \& {Fortney}}]{Williams06}
{Williams}, P.~K.~G., {Charbonneau}, D., {Cooper}, C.~S., {Showman}, A.~P., \&
  {Fortney}, J.~J. 2006, \apj, 649, 1020

\bibitem[{{Winn} {et~al.}(2007){Winn}, {Holman}, {Bakos}, {Pal}, {Johnson},
  {Williams}, {Shporer}, {Mazeh}, {Fernandez}, {Latham}, \& {Gillon}}]{Winn07b}
{Winn}, J.~N., {Holman}, M.~J., {Bakos}, G.~A., {Pal}, A., {Johnson}, J.~A.,
  {Williams}, P.~K.~G., {Shporer}, A., {Mazeh}, T., {Fernandez}, J., {Latham},
  D.~W., \& {Gillon}, M. 2007, \aj, 134, 1707

\bibitem[{{Yelle}(2004)}]{Yelle04}
{Yelle}, R.~V. 2004, Icarus, 170, 167

\end{thebibliography}

\clearpage

\end{document}